\documentclass{article}
\usepackage{epsfig}
\usepackage{graphicx}
\usepackage[centertags]{amsmath}
\usepackage{amssymb}
\usepackage[square,numbers,sort&compress]{natbib}

\begin{document}

\title{STIRAP-like transitions in a harmonically-modulated optical
  lattice}

\author{Benjamin P. Holder and Linda E. Reichl\\
Center for Complex Quantum Systems and Department of Physics,\\
The University of Texas at Austin, Austin, Texas 78712}

\date{February 26, 2007}

\maketitle

\begin{abstract}

We introduce a method for the coherent acceleration of atoms trapped
in an optical lattice, using the well-known model for stimulated
Raman adiabatic passage (STIRAP). Specifically, we show that small
harmonic modulations of the optical lattice amplitude, with
frequencies tuned to the eigenvalue spacings of three
``unperturbed'' eigenstates, reveals a three-state STIRAP subsystem.
We use this model to realize an experimentally achievable method for
transferring trapped atoms from stationary to motional eigenstates.

\end{abstract}


%
%
\section{\label{sec:1} Introduction}
Stimulated Raman adiabatic passage (STIRAP) is a method for achieving
coherent transitions between quantum states by applying two coupling fields
in a non-intuitive pulse sequence.  The frequencies of these fields are
tuned to the eigenvalue spacings between the ``initial'' and ``target''
states and a third ``intermediate'' state.  When the coupling of the target
and intermediate states precedes that of the initial and intermediate states
in time, population transfer from the initial to target state is achieved.
In the adiabatic limit, the transition is $100\%$ efficient and involves no
occupation of the intermediate state.  The use of STIRAP for atomic and
molecular systems was first demonstrated experimentally by Gaubatz and
coworkers \cite{gaubatz1988, gaubatz1990}, who achieved population transfer
between vibrational levels in a beam of sodium molecules. Further
references on the STIRAP transitions in atoms and molecules can be found in
the review by Vitanov {\em et al} \cite{vitanov2001}. In atom optics
experiments, STIRAP has been used for coherent momentum tranfer
\cite{marte_zoller_hall1991,goldner1994}, and velocity-selective coherent
population trapping for laser cooling of trapped atoms
\cite{esslinger1996,kulin1997}.  Extensions of STIRAP, with a particular
focus on the influence of quantum chaos, have been studied by Na and Reichl
\cite{na_reichl2004,na_reichl2005}.  Na {\em et al}
\cite{na_jung_reichl2006} have also use STIRAP to control the isomerization
transition of HOCl.  

As a theoretical model, STIRAP can be defined by the adiabatic
behavior of a three-level system, which in some basis $( | a
\rangle, | b \rangle, | c \rangle)$ is represented by the
Hamiltonian
\begin{figure}
  \centering
  \includegraphics[width=1\textwidth]{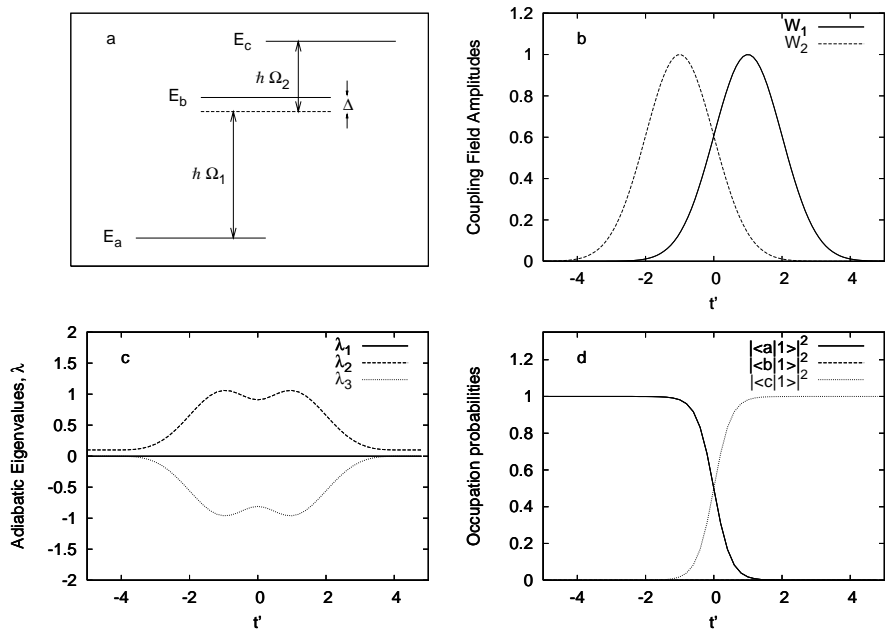}
  \caption{ The 3-level ``ladder'' STIRAP system (Eq. \ref{eqn:STIRAP}) with parameters
    $\Delta=0.05$, $W_1=W_2=1$ and $\hbar=2$.  Coupling fields are applied with
    frequencies equally detuned from the spacings of the unperturbed
    energy levels by $\Delta$ (a).  Adiabatic variation of amplitudes
    of the coupling fields (b) in the manner described by
    Eq. (\ref{eqn:stirap_conditions}) affects a transition of the
$|1 \rangle$ eigenvector between basis states $|a \rangle$ and $|b
\rangle$ (d).  The eigenvalue corresponding to this state remains
unchanged at zero throughout the transition (c)}
  \label{fig:stirap_3level}
\end{figure}
\begin{equation} \label{eqn:STIRAP}
H(t^{\prime}) = -\frac{\hbar}{2} \left( \begin{array}{ccc}
  0 & W_1(t^{\prime}) & 0 \\
  W_1(t^{\prime}) & -2 \Delta & W_2(t^{\prime}) \\
  0 & W_2(t^{\prime}) & 0 \\
\end{array} \right)\,,
\end{equation}
under the variation of the parameters $W_1$ and $W_2$, with $\Delta$ a
constant.  This model, which we refer to as the ``STIRAP model''
hereafter, was first introduced by Kuklinski {\em et al}
\cite{kuklinski1989}, following significant work by Hioe {\em et al}
\cite{hioe1983,oreg_hioe1984,carroll_hioe1988}, to succinctly describe
the experimental results of Gaubatz {\em et al}
\cite{gaubatz1988}. Starting from a system in which the pairs of
states $(|a \rangle, |b \rangle)$ and $(|b \rangle, |c \rangle)$ are
each dipole-coupled by monochromatic electric fields, the STIRAP model
is derived by applying the rotating-wave approximation and assuming an
equal detuning of the coupling frequencies $\Delta$ (see Figure
\ref{fig:stirap_3level}).  In that particular system the
$W_i(t^{\prime})$ are the Rabi oscillation frequencies corresponding
to the two couplings. The STIRAP transition, however, does not depend
on the physical system from which Eq. (\ref{eqn:STIRAP}) is
derived. In a novel application of this model by Eckert {\em et al}
\cite{eckert2004}, for example, the functions $W_i$ were related to
the spatial separations of three optical microtraps in order to induce
coherent transport of atoms between the ground states of the two
farthest separated traps.

The matrix in Eq. (\ref{eqn:STIRAP}) allows for a transition of the
type described above because of the existence of the eigenvector
\begin{equation} \label{eqn:stirap_eigenvector}
| 1 (t^{\prime})\rangle = \cos \theta(t^{\prime}) | a \rangle - \sin
\theta(t^{\prime}) | c\rangle \,,
\end{equation}
where $\tan \theta(t^{\prime}) \equiv
\frac{W_1(t^{\prime})}{W_2(t^{\prime})}$.  Under the conditions
\begin{equation} \label{eqn:stirap_conditions}
\lim_{t^{\prime} \rightarrow - \infty}
\frac{W_1(t^{\prime})}{W_2(t^{\prime})} \rightarrow 0 \quad {\rm
and} \quad \lim_{t^{\prime} \rightarrow + \infty}
\frac{W_2(t^{\prime})}{W_1(t^{\prime})} \rightarrow 0\,,
\end{equation}
the adiabatic evolution of state $| 1 \rangle$ is from $|a \rangle$ to $|c
\rangle$.  Thus, the transition is achieved by first coupling the upper two
levels and then coupling the lower two (in some continuous, e.g. Gaussian,
manner). Moreover, because of the form of the eigenstate $|1 \rangle$, the
state $| b \rangle$ remains unoccupied throughout the transition.

In this paper, we apply the method of STIRAP to the motional states of atoms
in an optical lattice.  As was first shown by Graham, Schlautmann and Zoller
\cite{gs_zoller1992}, the interaction of a single transition in an alkali
atom with a pair of counter-propagating lasers can be reduced to an
effective Hamiltonian for the center-of-mass motion of the atom in a cosine
potential. Modulation of the laser amplitudes and/or the introduction of
laser pairs with offset frequencies introduces a periodic time-dependence
(see Appendix A). Here we will analyze the ``two-resonance'' system
described by the effective Hamiltonian
\begin{equation} \label{eqn:tworesham}
\hat{H}_0(t) = \hat{p}^2 + \kappa_0 \left[ \cos\hat{x} + \cos
(\hat{x} - \omega_0 t) \right] \,,
\end{equation}
where each of the two cosine terms are produced by a pair of
counter-propagating lasers with $\kappa_0$ proportional to the
square of the laser amplitudes.  We show, using perturbation
analysis of an associated Floquet Hamiltonian, that small harmonic
modulations of the laser amplitudes can be used to affect a
STIRAP-like transition from a state localized in the stationary
cosine well into a state localized in the traveling cosine well.

In Section \ref{sec:2} we present an analysis of the
time-independent ``quantum pendulum'' system which reveals a
three-level STIRAP model in the regime of small perturbation.
Section \ref{sec:3} contains a slightly modified method in order to
obtain a STIRAP-like model for the time-dependent ``two-resonance''
system. In each case, adiabatic results are compared to numerical
evolution of the Schr\"odinger equation and transitions of nearly
$100\%$ effeciency are observed.  Concluding remarks are presented in
Section \ref{sec:4}.

\section{\label{sec:2} STIRAP transitions in the Quantum Pendulum}

Before analyzing the two-resonance effective Hamiltonian presented
in the introduction, we will consider STIRAP-like transitions
within the quantum pendulum system:
\begin{equation} \label{eqn:quantumpend}
\hat{H}_{pend} = \hat{p}^2 + \kappa_0 \cos \hat{x}\,.
\end{equation}
This is the simplest type of effective Hamiltonian for optical lattice
experiments, achieved with a single pair of counter-propagating lasers
with equal frequencies. The parameter $\kappa_0$ is proportional to
the square of the electric field amplitude.  Because of experimental
techniques which can limit momentum values to the integers (see
Appendix \ref{appendix:a}), the eigenstates can be considered
considered spatially periodic with period $2 \pi$. The position space
solutions to the eigenvalue equation $\hat{H}_{pend} | \chi_n \rangle
= E_n | \chi_n \rangle$ are the Mathieu functions $ \langle x | \chi_n
\rangle \; (n \in Z)$ \cite{abramowitz}, with $n$ even labeling
even-parity functions and $n$ odd labeling odd-parity functions.

To affect a transition in this system using STIRAP, we add a
time-periodic modulation of the lattice amplitude of the form
\begin{equation} \label{eqn:floquet_pert}
\lambda \hat{V}(t) = \lambda \cos \hat{x} \left[ \kappa_1 \cos
(\Omega_1 t) +
  \kappa_2 \cos (\Omega_2 t) \right]\,,
\end{equation}
where $\lambda$ is small, and $\Omega_1$ and $\Omega_2$ are
commensurate with $\frac{\Omega_1}{\Omega_2}= \frac{m_1}{m_2}$ ($m_i
\in Z$).  It is useful to write $\Omega_1$ and $\Omega_2$ in terms
of a common frequency $\omega$ such that $\Omega_1=m_1 \,\omega$ and
$\Omega_2=m_2 \,\omega$, where $\omega = \frac{2 \pi}{T}$ and $T$ is
the periodicity of the perturbation. This perturbation is achieved
experimentally by modulating the intensity of the
counter-propagating laser radiation about the $\kappa_0$ value,
meaning that the lasers' electric field amplitude $E(t)$ should take
the form
\begin{equation}
|E(t)|^2 \sim \kappa_0 + \lambda \left[ \kappa_1 \cos ( m_1 \omega
t) + \kappa_2 \cos (m_2 \omega t) \right]\,.
\end{equation}
For now, we will consider the coefficients $\kappa_1$ and $\kappa_2$
to have constant values.  Perturbation analysis will reveal that, for
small values of $\lambda$, there exists a $3$-state subsystem
identical to the STIRAP model, parameterized by these coefficients.  A
STIRAP-type transition can then be affected by adiabatic variation of
$\kappa_1$ and $\kappa_2$ in the manner described in Eq.
(\ref{eqn:stirap_conditions}).  The justification of this
time-parameterization of the $\kappa_i$, following a Floquet analysis
of the system where they are considered constant, is provided in
Appendix \ref{appendix:b}.

Having required that the frequencies $\Omega_1$ and $\Omega_2$ are
commensurate, we can analyze the dynamics of the full system
\begin{equation} \label{eqn:pend_schrod}
i \frac{\partial}{\partial t} | \psi(t) \rangle = \left[ \hat{H}_{pend}
+ \lambda \hat{V}(t) \right] | \psi (t) \rangle
\end{equation}
using Floquet theory, which we review briefly here. Any solution of
Eq. (\ref{eqn:pend_schrod}) can be written in the form
\begin{equation} \label{eqn:floquet_sol}
| \psi_{\alpha}(t) \rangle = {\rm e}^{-i \epsilon_{\alpha} t} |
  \phi_{\alpha}(t) \rangle \,,
\end{equation}
where the {\em Floquet eigenstate} $|\phi_{\alpha}(t) \rangle$ is periodic
in time with period $T$, and $\epsilon_{\alpha}$ is called the {\em Floquet
eigenvalue}. Plugging this solution into the Schr\"odinger equation we
arrive at the eigenvalue equation
\begin{equation} \label{eqn:floquetham_pend}
\hat{H}_F | \phi_{\alpha} \rangle \equiv \left[ \hat{H}_{pend} +
\lambda  \hat{V}(t) - i \frac{\partial}{\partial t} \right] |
\phi_{\alpha} \rangle = \epsilon_{\alpha} | \phi_{\alpha} \rangle
\,,
\end{equation}
where the {\em Floquet Hamiltonian} $\hat{H}_F$ is a Hermitian
operator in an extended Hilbert space which has time as a periodic
{\em coordinate} \cite{holder_reichl2005,sambe1973,dzyublik1991}.
Diagonalization of $\hat{H}_F$ in some appropriate basis in this
space yields the Floquet eigenstates and eigenvalues. The Floquet
eigenstates have an infinite multiplicity with respect to the
solutions of the Schr\"odinger equation in the sense that for any
Floquet eigenstate $|\phi \rangle$ with eigenvalue $\epsilon$, there
exists another solution $| \tilde{\phi} \rangle = {\rm e}^{i Q
\omega t} | \phi \rangle$ ($Q \in \mathbb{Z}$) with eigenvalue
$\tilde{\epsilon} = \epsilon + Q \omega$.  These two Floquet states
are, however, associated to the same physical state, i.e. $ |\psi
\rangle \equiv {\rm e}^{-i \epsilon t} | \phi \rangle = {\rm e}^{-i
\tilde{\epsilon} t} | \tilde{\phi} \rangle$.  The implication of
this fact is that the dynamics of the physical system can be
understood by considering only those Floquet states whose
eigenvalues appear in a single ``zone'' $\epsilon^{\star} \le
\epsilon < \epsilon^{\star} + \omega$, labeled by the constant
$\epsilon^{\star}$.

The STIRAP model system is derived by applying perturbation theory to
Eq. ({\ref{eqn:floquetham_pend}), where the two frequencies in
$\hat{V}(t)$ are chosen to ``couple" three pendulum eigenvalues, at a
particular value of $\kappa_0$, in the manner shown in Figure
\ref{fig:stirap_3level}a. Although, in general, the ratio of these
eigenvalue spacings is not rational, the equal detuning $\Delta$
allows for $\Omega_1$ and $\Omega_2$ to be chosen as commensurate.
More precisely, given three eigenvalues of the quantum pendulum $E_a <
E_b< E_c$, any pair of integers $(m_1,m_2)$ uniquely determines
$\Delta$ and $\omega$ via the coupled equations
\begin{equation} \label{eqn:Delta_omega}
\begin{split}
m_1 \, \omega &= E_b - E_a - \Delta \\
m_2 \, \omega &= E_c - E_b + \Delta\,.
\end{split}
\end{equation}
Eliminating $\omega$ we obtain an expression for $\Delta$
\begin{equation}
\Delta =  \frac{m_2 \, \omega_1 - m_1 \, \omega_2}{m_1 + m_2}\,,
\end{equation}
where we have defined $\omega_1 = E_b - E_a$ and $ \omega_2 = E_c -
E_b$. We can see that the integer vectors $\vec{m} \equiv ( m_1,
m_2)^{\rm T}$ which minimize $\Delta$ are those which satisfy
\begin{equation}
\vec{m} \cdot \vec{\nu} \approx 0 \quad {\rm with} \quad \vec{\nu}
\equiv (-\omega_2, \omega_1)^{\rm T}
\end{equation}
Therefore, the best choices for $\vec{m}$ are those for which $m_1 /
m_2$ are the best rational approximants of the ratio $w \equiv
\omega_1 / \omega_2$.

\begin{figure}
  \centering
  \includegraphics[width=1\textwidth]{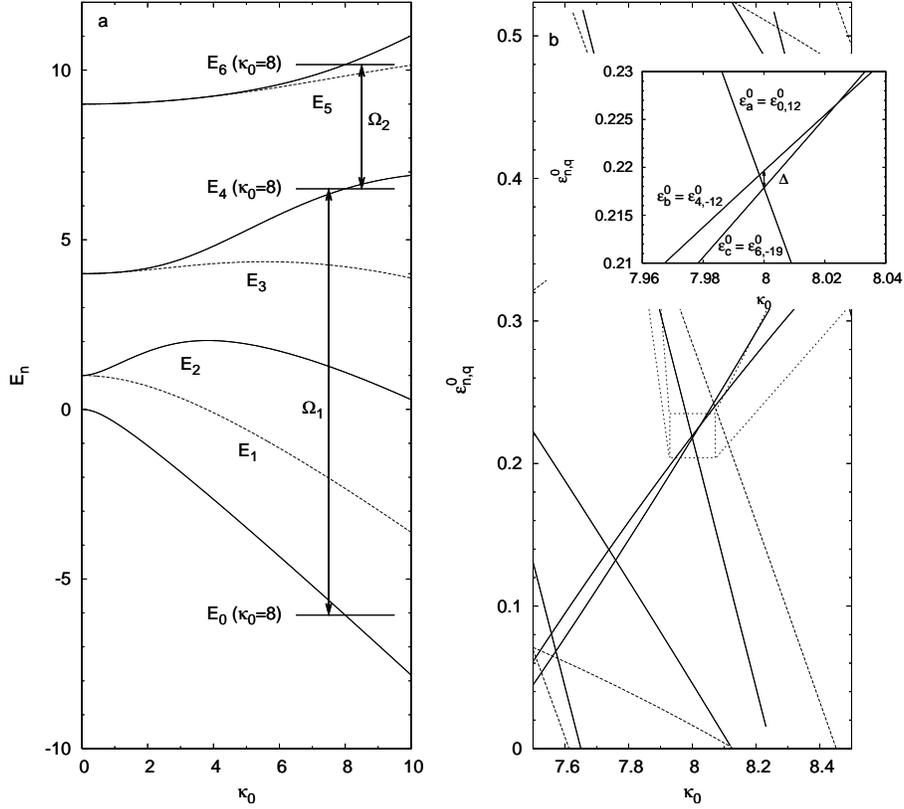}
  \caption{ The lowest seven eigenvalues of the quantum
    pendulum (a) as a function of the parameter $\kappa_0$ (solid lines
    are the eigenvalues of even-parity states, dashed odd).  The
    coupling of levels $E_0$-$E_4$ and $E_4$-$E_6$ at $\kappa_0=8$
    via the perturbation $\hat{V}(t)$ requires a Floquet treatment of
    the pendulum system.  The corresponding Floquet eigenvalues
    (for $\omega=0.5235$) are shown in (b). The degeneracy and
    near-degeneracy of the the three coupled eigenvalues can be seen
    in the enlarged (inset) view.}
  \label{fig:quantpend_stirap}
\end{figure}
Let us denote the unperturbed Floquet Hamiltonian ($\hat{H}_F$ when
$\lambda = 0$) as
\begin{equation}
\hat{H}^0_F \equiv \hat{H}_{pend} - i \frac{\partial}{\partial t}\,.
\end{equation}
In the extended Hilbert space, $\hat{H}^0_F$ has normalized
eigenvectors of the form
\begin{equation} \label{eqn:floquetpend_evecs}
\langle t| \chi_n ,q \rangle = \langle t | q \rangle | \chi_n
\rangle  = \frac{1}{\sqrt{T}} \,{\rm e}^{i q \omega t} |\chi_n
\rangle \quad,\quad q \in \mathbb{Z}\,.
\end{equation}
The corresponding eigenvalues are $\epsilon^0_{n,q} = E_n + q
\omega$. We are interested in the dynamics of an initial population of
atoms localized in the pendulum state $| \chi_a \rangle$. With
$\lambda=0$, state $|\chi_a \rangle$ is represented in a particular
zone of Floquet eigenvalues of $\hat{H}_F^0$ by a Floquet eigenstate
$| \chi_a, q_a \rangle$ with eigenvalue $\epsilon_{a,q_a}$ in that
zone.  The coupling frequencies $\Omega_1$ and $\Omega_2$ have been
chosen in Eq. (\ref{eqn:Delta_omega}) such that $\epsilon^0_{c,q_c}$,
the Floquet eigenvalue in that zone corresponding to the physical
state $| \chi_c \rangle$, is equal to $\epsilon^0_{a,q_a}$ and the
eigenvalue $\epsilon^0_{b,q_b}$ is offset from this value by
$\Delta$. The degeneracy of these two eigenvalues and the
near-degeneracy of the third requires that any perturbation analysis
must be performed in the degenerate form
\cite{holder_reichl2005}. Therefore, we expand the extended Hilbert
space state $| \phi \rangle$ in powers of the small parameter
$\lambda$
\begin{equation}
| \phi \rangle = | \phi^{(0)} \rangle + \lambda | \phi^{(1)} \rangle +
  \lambda^2 | \phi^{(2)} \rangle + \ldots
\end{equation}
and take the zeroth-order state to be a superposition of the three
degenerate or near-degenerate eigenstates of $\hat{H}_F^0$:
\begin{equation}
|\phi^{(0)} \rangle = C_a |\chi_a,q_a \rangle + C_b |\chi_b,
q_b\rangle + C_c |\chi_c,q_c \rangle \,.
\end{equation}
The eigenvalue is likewise expanded in orders of the small parameter
\begin{equation}
\epsilon = \epsilon^{(0)} + \lambda \epsilon^{(1)} + \lambda^2
\epsilon^{(2)} + \ldots\,.
\end{equation}
For brevity of notation, we write the unperturbed eigenstates $|a
\rangle \equiv | \chi_{a}, q_a \rangle$, $|b\rangle \equiv |\chi_b,
q_b \rangle$ and $|c \rangle \equiv |\chi_c,q_c \rangle$, with
associated eigenvalues $\epsilon^0_a \equiv \epsilon^0_{a,q_a}$,
$\epsilon^0_b \equiv \epsilon^0_{b,q_b}$ and $\epsilon^0_c \equiv
\epsilon^0_{c,q_c}$, respectively.

\begin{figure}
  \centering
  \includegraphics[width=1\textwidth]{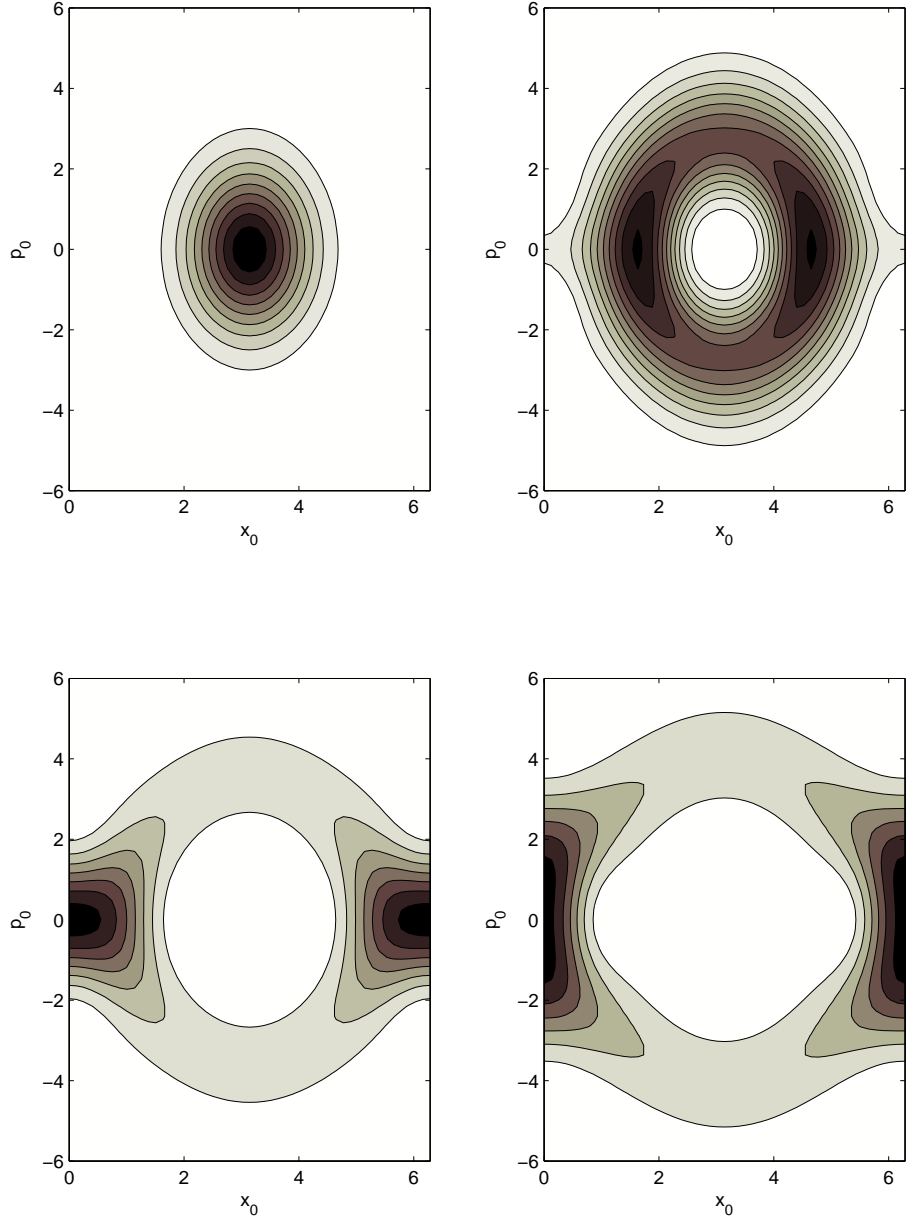}
  \caption{The Husimi functions of the first four even-parity
    eigenstates of the quantum pendulum on classical phase space $(x_0,p_0)$.
    The states are, clockwise from top left: $| \chi_0 \rangle$,$| \chi_2 \rangle$,$|\chi_6 \rangle$,$| \chi_4 \rangle$.   }
  \label{fig:pend_husimis}
\end{figure}
Retaining terms up to first-order in $\lambda$, we obtain an
isolated subsystem of Equation (\ref{eqn:floquetham_pend}):
\begin{equation} \label{eqn:firstorder_pend}
\left( \begin{array}{ccc} \epsilon^0_a + \lambda V_{a,a} & \lambda
V_{a,b} & \lambda V_{a,c} \\
\lambda V_{b,a} & \epsilon^0_b + \lambda V_{b,b} & \lambda V_{b,c} \\
\lambda V_{c,a} & \lambda V_{c,b} & \epsilon^0_c + \lambda V_{c,c}
\end{array} \right)
\left( \begin{array}{c}
C_a \\
C_b \\
C_c
\end{array} \right)
= \epsilon^{\prime} \left( \begin{array}{c}
C_a \\
C_b \\
C_c
\end{array} \right)\,,
\end{equation}
where $\epsilon^{\prime} \equiv \epsilon^0_a + \lambda
\epsilon^{(1)}$ and we have assumed that $\Delta$ is of order
$\lambda$. The matrix elements of the perturbation are determined in
the following way:
\begin{align} \label{eqn:matrix_elem_pend}
V_{i,j} &= \langle \langle \chi_i,q_i | \hat{V} | \chi_j, q_j
\rangle \rangle \equiv \int_{-\pi / \omega}^{\pi / \omega} \langle
q_i | t \rangle \langle t | q_j \rangle \langle \chi_i | \cos
\hat{x} | \chi_j \rangle \nonumber \\
&\hspace{2in} \times \left[\kappa_1 \cos (m_1 \omega t)
+ \kappa_2 \cos (m_2 \omega t) \right] dt \nonumber \\
&= \frac{\langle \chi_i | \cos \hat{x} | \chi_j
\rangle}{2} \\
&\quad \times \left[\kappa_1 \delta_{q_j, q_i+m_1} + \kappa_1
  \delta_{q_j, q_i-m_1} + \kappa_2 \delta_{q_j, q_i+m_2} +
  \kappa_2 \delta_{q_j, q_i-m_2} \right] \,.
\nonumber
\end{align}
The selection of $\Omega_1$ and $\Omega_2$ guarantees that the $q$
indices for these states satisfy
\begin{equation}
q_b-q_a=m_1 \quad {\rm and} \quad q_c-q_b=m_2 \,.
\end{equation}
Therefore, the only non-zero matrix elements of the perturbation are
$V_{a,b}=V_{b,a}$ and $V_{b,c}=V_{c,b}$.  Further, we can write
\begin{equation}
V_{a,b} = V_{b,a} = \kappa_1 \, v_{a,b} \quad {\rm and} \quad
V_{b,c} = V_{c,b}= \kappa_2 \, v_{b,c}
\end{equation}
where the numerical values of
\begin{equation}
v_{i,j}=\frac{\langle \chi_i | \cos \hat{x} | \chi_j \rangle}{2}
\end{equation}
are calculated using the Mathieu functions. Subtracting
$\epsilon^0_a \,(C_a,C_b,C_c)^{\rm T}$ from both sides of Eq.
(\ref{eqn:firstorder_pend}) and redefining $\epsilon^{\prime} -
\epsilon_a^0 \rightarrow \epsilon^{\prime}$, we arrive at
\begin{equation} \label{eqn:STIRAP_pend}
\left( \begin{array}{ccc}
0 & \lambda \, \kappa_1 \, v_{a,b} & 0 \\
\lambda \, \kappa_1 \, v_{b,a} & \Delta & \lambda \, \kappa_2 \, v_{b,c} \\
0 & \lambda \, \kappa_2 \, v_{b,c} & 0
\end{array} \right)
\left( \begin{array}{c}
C_{a} \\
C_{b} \\
C_{c}
\end{array} \right)
=
\epsilon^{\prime}
\left( \begin{array}{c}
C_{a} \\
C_{b} \\
C_{c}
\end{array} \right)\,,
\end{equation}
which is equivalent to the STIRAP model system.  Thus, in the limit of
small $\lambda$ (and $\Delta$), the parameters $\kappa_1$ and
$\kappa_2$ can be adiabatically varied in the manner described in the
introduction to affect a transition between the unperturbed states $|
a \rangle$ and $| c \rangle$.

We now provide a concrete example on which to demonstrate the
analysis. Figure \ref{fig:quantpend_stirap}a shows the energies of
the first few eigenstates of $\hat{H}_{pend}$ as a function of
$\kappa_0$.  Husimi representations \cite{husimi, holder_reichl2005}
of the even eigenstates, at $\kappa_0=8$, are shown in Figure
\ref{fig:pend_husimis}. At this value of $\kappa_0$, we choose
energies $E_a=E_0$, $E_b=E_4$ and $E_c=E_6$ for coupling. These
energy levels have spacings $\omega_1 = E_b-E_a = 12.5668395$ and
$\omega_2=E_c-E_b = 3.6630472$ with ratio
\begin{equation}
w = 3.43070639... = [3,2,3,9,...] = 3 + \frac{1}{2+ \frac{1}{3 +
\frac{1}{9 + \cdots}}}\,.
\end{equation}
Thus, the best rational approximates of $w$, found by truncating the
continued fraction, are $\{\frac{3}{1}, \frac{7}{2}, \frac{24}{7},
\frac{223}{65},...\}$. Using the third approximation, the modulation
frequencies shown in the example have been chosen to be $\Omega_1 = 24
\omega$ and $\Omega_2 = 7 \omega$, giving $\omega=0.5235$ and $\Delta=1.766
\times 10^{-3}$.

In Figure \ref{fig:quantpend_stirap}b, the Floquet eigenvalues of
$H^0_F$ are shown in the zone $\epsilon^{\star}=0$. In the inset
figure, one can see that $\epsilon^0_a=\epsilon^0_{0,12}$ and
$\epsilon^0_c=\epsilon^0_{6,-19}$ are equal at $\kappa_0=8$ and
$\epsilon^0_b=\epsilon^0_{4,-12}$ is offset by $\Delta$.  Using the
Mathieu functions, the perturbation matrix elements are calculated
to be
\begin{align}
V_{a,b} &= V_{b,a} = \kappa_1 \, v_{a,b} = - 1.16 \times 10^{-2} \kappa_1\\
V_{b,c} &= V_{c,b}= \kappa_2 \, v_{b,c} = 2.50 \times 10^{-1}
\kappa_2 \,.
\end{align}
\begin{figure}
  \centering
  \includegraphics[width=1\textwidth]{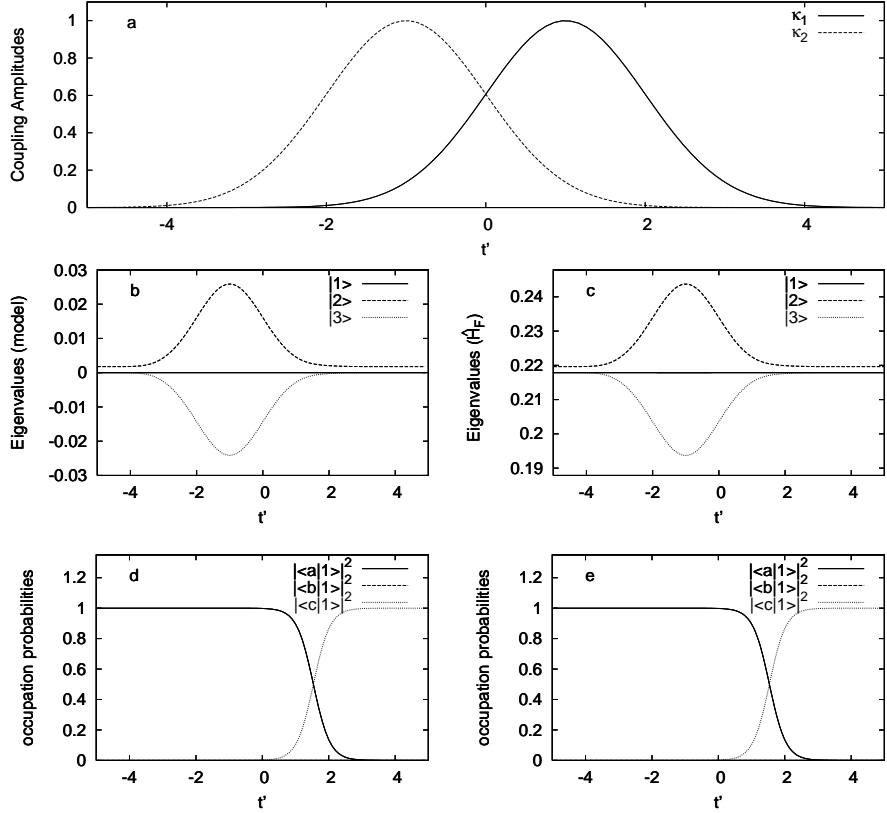}
  \caption{Adiabatic evolution of the pendulum system under STIRAP
coupling, comparing the dynamics of the model in Eq.
(\ref{eqn:STIRAP_pend}) (b and d) to that of the full Floquet
Hamiltonian in Eq. (\ref{eqn:pend_schrod}) (c and e) for
$\lambda=0.1$. The coupling field $\kappa_2$ is seen to have a
greater effect on the eigenvalues at $t^{\prime}=-1$ than that of
$\kappa_1$ at $t^{\prime}=1$, since $v_{b,c}>>v_{a,b}$.  This
asymmetry also shifts the transition of the $|1\rangle$ eigenstate
from pendulum states $|a \rangle$ to $|c\rangle$ to a
later time (d and e).}
  \label{fig:pend-adiab_3st_n_full}
\end{figure}
To accomplish a STIRAP transition from $|a \rangle$ to $|c \rangle$,
we give $\kappa_1$ and $\kappa_2$ Gaussian dependence on an adiabatic
time parameter $t^{\prime}$ (see Figure
\ref{fig:pend-adiab_3st_n_full}a):
\begin{equation} \label{eqn:kappai_gauss}
\kappa_i(t^{\prime}) = \exp \left[ - \frac{(t^{\prime} -
t_i)^2}{2 \sigma_i^2} \right] \,.
\end{equation}
The conditions of Eq. (\ref{eqn:stirap_conditions}) are satisfied by
setting $ t_1 = - t_2 = 1.0$ and
$\sigma_1=\sigma_2=1.0$. Figure \ref{fig:pend-adiab_3st_n_full}
shows good agreement between the adiabatic dynamics of the model
system in Eq. (\ref{eqn:STIRAP_pend})
(\ref{fig:pend-adiab_3st_n_full}b and d), and that of the full
Floquet Hamiltonian (\ref{fig:pend-adiab_3st_n_full}c and e).

The implementation of this transition in an experimental system (or
the numerical evolution of the Schr\"odinger equation) is not
dependent on the time-periodicity which we have required thus
far. Floquet analysis has proven an essential theoretical tool for
revealing the existence of the STIRAP model, but the method has
introduced no upper limit on the integers $m_1$ and $m_2$ whose ratio
approximates $\omega_1/\omega_2$.  Therefore we may choose the
coupling frequencies to be resonant ($\Omega_1 = \omega_1$ and
$\Omega_2 = \omega_2$) to any desired accuracy.  The results for the
numerical evolution of the effective Schr\"odinger equation
(\ref{eqn:pend_schrod}) are shown in Figure
\ref{fig:evolve_pend_w-hus}a, for the case of both resonant and
near-resonant coupling ($\Delta=1.77 \times 10^{-3}$) with
$\lambda=0.1$.  The evolution was performed over a set time period
$[0,t_{tot}]$ with initial condition $|\langle \chi_a | \psi
\rangle|^2=1$, and Gaussian parameters for the coupling amplitudes
$\kappa_1$ and $\kappa_2$ of $\sigma_1=\sigma_2=0.1 t_{tot}$, $t_1=0.6
t_{tot}$ and $t_2=0.4 t_{tot}$. It is seen that resonant coupling
provides a more rapid approach to the adiabatic behavior.  In Figure
\ref{fig:evolve_pend_w-hus}b, good agreement is seen between the
resonant evolution of the full effective Schr\"odinger equation and
the adiabatic predictions of Figure \ref{fig:pend-adiab_3st_n_full}.

\begin{figure}
  \centering
  \includegraphics[width=0.8\textwidth]{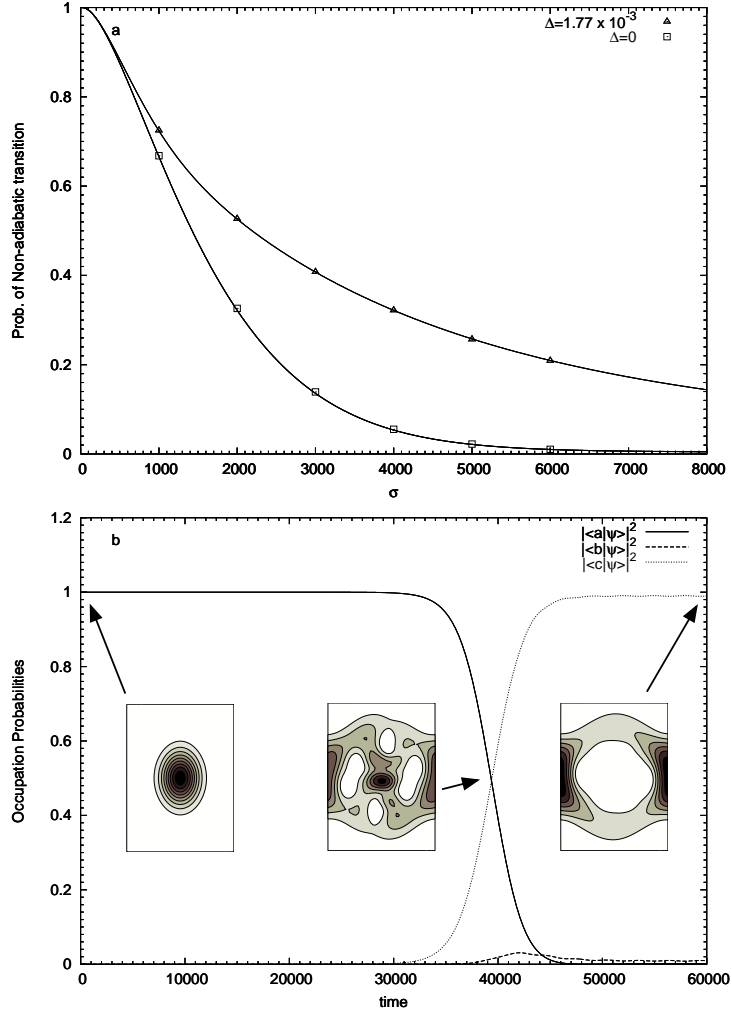}
  \caption{The numerical evolution of a state initially localized in
the $|\chi_a \rangle$ pendulum state, under STIRAP coupling of the
energies. The probability of a non-adiabatic transition (transition to
any state other than $| c \rangle$) is plotted versus the width of the
gaussian field $\sigma=\sigma_1=\sigma_2$ (a), for resonant
(triangles) and near-resonant coupling (squares) with $\lambda=0.1$.
The solid lines are the values predicted by evolution of the
three-state model (see Appendix \ref{appendix:b} for a justification
of this use of the model).  Figure (b) shows the numerical evolution,
under the effective Schr\"odinger equation, of a state classically
localized within the pendulum well to one localized on the separatrix
(axes on inset Husimi functions are the same as in Fig.
\ref{fig:pend_husimis}). This evolution corresponds to the point at
$\sigma=6000$ with resonant ($\Delta=0$) coupling in (a).}
  \label{fig:evolve_pend_w-hus}
\end{figure}

\section{\label{sec:3} STIRAP transitions from stationary to moving
atoms}

We now consider the case in which the ``unperturbed'' Hamiltonian has
the form of Eq. (\ref{eqn:tworesham}), which consists classically of a
stationary cosine wave and a cosine wave that travels through phase
space with a speed $\omega_0$. Our goal is to cause a coherent
transition of an entire cloud of trapped atoms from a state localized
in the stationary wave (in the sense of its Husimi distribution) into
a state localized in the traveling wave, so that the entire collection
of atoms changes velocity from $v = 0$ to $v = \omega_0$.

Our approach is analogous to that of the previous section. We apply
perturbation theory to a Floquet eigensystem of the form
\begin{equation} \label{eqn:twores_pert_floqham}
\hat{H}_F(t) | \phi (t) \rangle = \left[ \hat{H}^0_F(t) + \lambda
\hat{V}(t) \right] | \phi (t) \rangle = \epsilon | \phi (t)
\rangle \,,
\end{equation}
where $\hat{H}_F$ is periodic in time with period $T = 2 \pi /
\omega$ and the perturbation operator $\hat{V}$ has the same form as
in Eq. (\ref{eqn:floquet_pert}). Again we find that, in the limit of
small $\lambda$, there exists an isolated three-level subsystem in
which a STIRAP-type transition between eigenstates of $\hat{H}^0_F$
can be induced.  The construction of $\hat{H}_F$, however, differs
significantly from the previous section because of the explicit
time-dependence of the two-resonance Hamiltonian. In the pendulum
analysis the frequencies $\Omega_1$ and $\Omega_2$ were chosen to
couple the energies of pendulum eigenstates.  Here, these
frequencies are chosen to couple the eigenvalues of the
two-resonance Floquet Hamiltonian,
\begin{equation} \label{eqn:twores_floquetham}
\hat{H}^0_F(t) = \hat{p}^2 + \kappa_0 \left[ \cos \hat{x} + \cos (
\hat{x} - \omega_0 t ) \right] - i \frac{\partial}{\partial t}\,,
\end{equation}
within a particular zone.  Selection of coupling frequencies such
that they and $\omega_0$ are commensurate allows for Floquet
analysis of the full system, but requires that the eigenvectors of
$\hat{H}^0_F$ be translated from their natural Hilbert space,
containing functions periodic in time with period $\bar{T} \equiv 2
\pi / \omega_0$, into the space containing $T$-periodic functions of
time. The relevant Floquet eigenvalues associated to the
eigenvectors in this latter space take near-degenerate values and
perturbation analysis leads to similar results as the previous
section. [It should also be noted that although we call the
two-resonance system in Eq. (\ref{eqn:twores_floquetham}) an
``unperturbed'' Hamiltonian, it is not analytically solvable.
Perturbation theory will be a useful tool to demonstrate the
existence of a STIRAP-like model for this system, but the
eigenvectors of $\hat{H}^0_F$ and all related quantities (e.g. the
matrix elements of the three-level system) must be determined
numerically.]

We begin by constructing Floquet eigenvectors $|
\bar{\phi}^{\,0}_{\alpha} \rangle$ of $\hat{H}^0_F$ (the ``overbar''
will be used to indicate that these vectors belong to the Hilbert
space $\bar{\mathcal{H}}$, defined below). We will assume that the
parameters $\kappa_0$ and $\omega_0$ have constant values, which may
be set arbitrarily. The only limitation on this choice is that,
given $\omega_0$, $\kappa_0$ should be chosen such that the set of
eigenvectors with eigenvalues in a particular zone contains one
state localized purely in the stationary cosine wave and one state
localized in the traveling wave (i.e. a $\kappa_0$ value far from
avoided crossings involving the eigenvalues of these states). The
eigenvectors $| \bar{\phi}^{\,0}_{\alpha} \rangle$ lie in the
extended Hilbert space $\bar{\mathcal{H}} \equiv \Theta \otimes
\bar{\mathcal{T}}$, where $\Theta$ is the space of all $2
\pi$-periodic, square-normalizable position-space functions and
$\bar{\mathcal{T}}$ is the space of all $\bar{T}$-periodic,
square-normalizable functions of time. We select the complete set of
momentum eigenstates $| n \rangle$ (see Appendix A) as a basis in
$\Theta$ and the analogous eigenstates $| q \rangle$ as a basis in
$\bar{\mathcal{T}}$, yielding normalized basis vectors in the
extended space which can be written
\begin{equation}
\langle x,t | n,q \rangle = \langle x|n \rangle \langle t | q
\rangle = \frac{1}{\sqrt{2 \pi \bar{T}}} {\rm e}^{i n x} {\rm e}^{i
q \omega_0 t}\,,
\end{equation}
with $n,q \in \mathbb{Z}$.  The eigenstates $| \bar{\phi}^{\,0}_{\alpha}
\rangle$ can then be written
\begin{equation}\label{eqn:barphi0}
| \bar{\phi}^{\,0}_{\alpha} (t) \rangle = \sum_{n,q}
\frac{1}{\sqrt{\bar{T}}} {\rm e}^{i q \omega_0 t} \langle n,q |
\bar{\phi}^{\,0}_{\alpha} \rangle \, | n \rangle\,,
\end{equation}
where $| \bar{\phi}^{\,0}_{\alpha} (t) \rangle = \langle t |
\bar{\phi}^{\,0}_{\alpha} \rangle$ and the coefficients $\langle n,q |
\bar{\phi}^{\,0}_{\alpha}\rangle$ are determined by diagonalization of
$\hat{H}^0_F$ in $\bar{\mathcal{H}}$.

We select a zone $\epsilon^{\star} \le \bar{\epsilon}^{\,0}_{\alpha}
< \epsilon^{\star} + \omega_0$ within which to perform a coupling of
the eigenvalues $\bar{\epsilon}^{\,0}_{\alpha}$ of $\hat{H}^0_F$.
Two of these eigenvalues, denoted $\bar{\epsilon}^{\,0}_a$ and
$\bar{\epsilon}^{\,0}_c$, are those of the states localized in the
stationary and traveling waves, respectively.  A third eigenvalue
$\bar{\epsilon}^{\,0}_b$ is chosen with the restriction that the
corresponding eigenvector is localized ``nearby'' in phase space
(the matrix element of $\cos \hat{x}$ between this and the other two
vectors should not be vanishingly small). As before, the coupling
frequencies $\Omega_1$ and $\Omega_2$ must be chosen to be
commensurate.  In this case, however, analogous equations to Eqs.
(\ref{eqn:Delta_omega}) cannot be solved simultaneously with the
requirement that $\omega_0$ is likewise commensurate:
\begin{equation} \label{eqn:omega0}
\omega_0=m_0 \, \omega \;\; (m_0 \in \mathbb{Z}).
\end{equation}
Therefore, in the following, we will relax the constant detuning
requirement and allow for two independent detunings defined by the
equations
\begin{equation} \label{eqn:Deltas_omega}
\begin{split}
\Omega_1&=m_1 \, \omega = \left( \bar{\epsilon}^{\,0}_b - 
\bar{\epsilon}^{\,0}_a \right) - \Delta_1 \\
\Omega_2&=m_2 \, \omega = \left( \bar{\epsilon}^{\,0}_c -
\bar{\epsilon}^{\,0}_b \right) + \Delta_2 \,.
\end{split}
\end{equation}
Given any integer vector $\vec{m} \equiv (m_0,m_1,m_2)^{\rm T}$,
Eqs. (\ref{eqn:omega0}) and (\ref{eqn:Deltas_omega}) can be
solved for $(\Delta_1, \Delta_2, \omega)$.  Eliminating $\omega$ and
defining $\omega_1 \equiv
\bar{\epsilon}^{\,0}_b-\bar{\epsilon}^{\,0}_a$ and $\omega_2 \equiv
\bar{\epsilon}^{\,0}_c-\bar{\epsilon}^{\,0}_b$, we obtain
\begin{equation} \label{eqn:Deltas_omega2}
\begin{split}
\Delta_1 &= \frac{ m_0 \, \omega_1\ - m_1 \, \omega_0}{m_0}\\
\Delta_2 &= \frac{ m_2 \, \omega_0 - m_0 \, \omega_2 }{m_0} \,.
\end{split}
\end{equation}
Thus, we see that the integer vectors $\vec{m}$ which simultaneously
minimize the two detunings will be those closest to the vector
perpendicular to the plane defined by $\vec{\nu}^{(1)} \equiv
(\omega_1, - \omega_0, 0)^{\rm T}$ and $\vec{\nu}^{(2)} \equiv
(-\omega_2, 0, \omega_0)^{\rm T}$.  This perpendicular vector is of
course $\vec{n} = (\omega_0, \omega_1, \omega_2)^{\rm T}$, and the
problem of minimizing the detunings is reduced to finding the best
integer approximate of $\vec{n}$ or, equivalently, finding the
simultaneous pair of rational approximants for $(\omega_1/\omega_0,
\omega_2/\omega_0)$. 

In the context of the three-level model system presented in the
introduction, non-equal detuning of the coupling frequencies leads to a
Hamiltonian of the form
\begin{equation}\label{eqn:STIRAP_detunings}
H = -\frac{\hbar}{2} \left( \begin{array}{ccc}
  0 & W_1 & 0 \\
  W_1 & -2 \Delta_1 & W_2 \\
  0 & W_2 & \Delta_2 - \Delta_1 \\
\end{array} \right)\,.
\end{equation}
It was recognized by Kuklinski {\em et al} \cite{kuklinski1989} that
this system could allow for a STIRAP transition, despite the absence
of an analytical result analogous to Eqs.
(\ref{eqn:stirap_eigenvector}) and (\ref{eqn:stirap_conditions}), as
long as the condition $\sqrt{W_1^2 + W_2^2} >> | \Delta_2 -
\Delta_1|$ is satisfied.  In order to satisfy this requirement, and
that of small perturbations, we will seek integer vectors $\vec{m}$
which provide detunings $|\Delta_2 - \Delta_1| << \Delta_1 << 1$.

Since the full, perturbed system $\hat{H}_F(t)$ is periodic in time with period
$T \equiv 2 \pi/\omega$, we must determine the eigenstates of the
unperturbed Floquet Hamiltonian in the extended Hilbert space
$\mathcal{H} \equiv \Theta \otimes \mathcal{T}$, where $\mathcal{T}$
is the Hilbert space of $T$-periodic functions. These eigenstates
$|\phi^0_{\alpha} \rangle$ can be expanded in $\mathcal{H}$ as
\begin{equation} \label{eqn:phi0}
|\phi^0_{\alpha}(t) \rangle = \sum_{n,q} \frac{1}{\sqrt{T}} \; {\rm
e}^{i q \omega t} \langle n, q |\phi^0_{\alpha} \rangle
  \; | n \rangle \,,
\end{equation}
where the $q$-eigenvectors now have the time-periodicity of
$\mathcal{T}$. Since the Schr\"odinger equation for the unperturbed
system
\begin{equation}
i \frac{\partial}{\partial t} | \psi (t)\rangle = \left\{ \hat{p}^2
+ \kappa_0 \left[ \cos \hat{x} + \cos ( \hat{x} - \omega_0 t)
\right] \right\} | \psi (t)\rangle
\end{equation}
can be viewed as time-periodic with either period $\bar{T}$ or $T=m_0
\bar{T}$, a physical solution $| \psi_{\alpha}(t) \rangle$ can be
written, using Eq. (\ref{eqn:floquet_sol}), in terms of a Floquet
state with either periodicity.  Equating these two representations, we
obtain a relationship between the Floquet eigenstates of $\hat{H}^0_F$
in spaces $\bar{\mathcal{H}}$ and $\mathcal{H}$:
\begin{equation}
\exp \left[-i \bar{\epsilon}^{\,0}_{\alpha} t \right]\, |
\bar{\phi}^{\,0}_{\alpha}(t) \rangle = A \,\exp \left[ -i
\epsilon^0_{\alpha} t \right] \, |\phi^0_{\alpha}(t)\rangle \,,
\end{equation}
where $\epsilon^0_{\alpha}$ is the eigenvalue associated to $|
\phi^0_{\alpha} (t) \rangle$ and $A$ is a proportionality constant.
Equating coefficients of the momentum eigenstate $| n \rangle$ in
Eqs. (\ref{eqn:barphi0}) and (\ref{eqn:phi0}), we find
\begin{equation} \label{eqn:phi_phibar}
\sum_q \exp[i q m_0 \omega t] \langle n,q |
\bar{\phi}^{\,0}_{\alpha} \rangle = \sum_{q^{\prime}} A^{\prime}
\,\exp \left[ i \left(q^{\prime} \omega - \epsilon^0_{\alpha} +
\bar{\epsilon}^{\,0}_{\alpha} \right) t \right] \langle n,q^{\prime}
| \phi^0_{\alpha} \rangle \,,
\end{equation}
where $A^{\prime}$ is again a constant.  A non-trivial solution to
this equation requires that the eigenvalues satisfy
$\epsilon^0_{\alpha} - \bar{\epsilon}^{\,0}_{\alpha} = Q \, \omega$,
where $Q$ is an integer. We see, then, that associated to each
Floquet eigenstate in $\hat{\mathcal{H}}$ is a family
of eigenstates in $\mathcal{H}$:
\begin{equation}
\left\{ | \phi^0_{\alpha,Q} \rangle , \epsilon^0_{\alpha,Q} \right\}
\quad  Q \in \mathbb{Z} \,,
\end{equation}
with $\epsilon^0_{\alpha,Q} = \bar{\epsilon}^{\,0}_{\alpha} + Q
\omega$. Selecting a particular value of $Q$, Equation
(\ref{eqn:phi_phibar}) becomes
\begin{equation}
\sum_q \exp \left[i q m_0 \omega t \right] \langle n,q |
\bar{\phi}^{\,0}_{\alpha} \rangle = \sum_{q^{\prime}} A^{\prime}\,
\exp \left[ i (q^{\prime}
  - Q ) \omega t \right] \langle n, q^{\prime} | \phi^0_{\alpha, Q} \rangle \,.
\end{equation}
Equating coefficients of the exponentials, we find
\begin{equation} \label{eqn:barredstates}
\langle n,q | \phi^0_{\alpha,Q} \rangle = \left\{
\begin{array}{cc}
\langle n, \frac{q-Q}{m_0}| \bar{\phi}^{\,0}_{\alpha} \rangle & {\rm
when}\;
\frac{q - Q}{m_0} \in Z \\
& \\
0 & {\rm otherwise}\,,
\end{array} \right.
\end{equation}
where we have set $A^{\prime}=1$ under normalization.  Therefore we
see that the unperturbed eigenstates in the space $\mathcal{H}$ have
non-zero coefficients $\langle n,q | \phi^0_{\alpha,Q} \rangle$ only
at $m_0$-separated values of $q$, with an offset of $Q$ from $q=0$.

Within a particular zone, we denote the unbarred eigenvalues
corresponding to $\left\{\bar{\epsilon}^{\,0}_a,
\bar{\epsilon}^{\,0}_b, \bar{\epsilon}^{\,0}_c\right\}$ as
$\left\{\epsilon^0_{a,Q_a}, \epsilon^0_{b,Q_b},
\epsilon^0_{c,Q_c}\right\}$, with values related by
\begin{equation} \label{eqn:deltas}
\begin{split}
\epsilon^0_{b,Q_b}-\epsilon^0_{a,Q_a} &= \Delta_1\\
\epsilon^0_{b,Q_b}-\epsilon^0_{c,Q_c} &= \Delta_2\,
\end{split}
\end{equation}
and corresponding eigenstates $\left\{| \phi^0_{a,Q_a} \rangle, |
\phi^0_{b,Q_b} \rangle,| \phi^0_{c,Q_c} \rangle\right\}$. The
$Q$-indices of these states are related by $Q_a-Q_b=m_1$ and
$Q_b-Q_c=m_2$.

Perturbation analysis of Eq. (\ref{eqn:twores_pert_floqham}) is now
performed by expanding the eigenstate $|\phi\rangle$ and eigenvalue
$\epsilon$ in powers of $\lambda$.  Assuming that $| \phi^0_{a,Q_a}
\rangle$ is initially occupied with probability one, and taking into
account the near-degeneracies of Eq. (\ref{eqn:deltas}), the
zeroth-order term in the expansion of the perturbed eigenstate is
chosen to be of the form
\begin{equation} \label{eqn:twores_phi0}
|\phi^{(0)} \rangle  = C_a | \phi^0_{a,Q_a} \rangle + C_b |
\phi^0_{b,Q_b} \rangle + C_c | \phi^0_{c,Q_c} \rangle \,.
\end{equation}
Retaining terms up to first order in $\lambda$ and making the
assumption that $\Delta_1$ and $\Delta_2$ are of order $\lambda$, we
obtain
\begin{equation} \label{eqn:firstorder_twores}
\left( \begin{array}{ccc} \epsilon^0_a + \lambda V_{a,a} &
\lambda V_{a,b} & \lambda V_{a,c} \\
\lambda V_{b,a} & \epsilon^0_b +\lambda V_{b,b} & \lambda V_{b,c}\\
\lambda V_{c,a} & \lambda V_{c,b} & \epsilon^0_c + \lambda V_{c,c}
\end{array} \right)
\left( \begin{array}c
C_a \\
C_b \\
C_c
\end{array} \right)
= \epsilon^{\prime} \left( \begin{array}c
C_a \\
C_b \\
C_c
\end{array} \right)\,,
\end{equation}
where $\epsilon^{\prime}\equiv \epsilon^0_{a} + \lambda
\epsilon^{(1)}$ and the matrix elements are calculated, defining $|
i \rangle \equiv | \phi_{i,Q_i} \rangle$, as follows
\begin{align}
V_{i,j} &\equiv \langle \langle i |\hat{V} | j \rangle \rangle =
\sum_{n,q,n^{\prime},q^{\prime}} \left[\langle i | n,q \rangle
\langle n | \cos \hat{x} | n^{\prime} \rangle \langle
n^{\prime}, q^{\prime}| j \rangle \right] \\
&\quad \times \left[ \kappa_1 \langle q | \cos (m_1 \omega \hat{t})
| q^{\prime} \rangle + \kappa_2 \langle q | \cos (m_2 \omega \hat{t})
| q^{\prime} \rangle \right] \nonumber \\
&= \frac{1}{2} \sum_{n,q,n^{\prime}} \kappa_1 \langle n | \cos
\hat{x} | n^{\prime} \rangle \left[ \langle i | n,q \rangle \langle
n^{\prime},q + m_1| j \rangle + \langle i | n,q
\rangle \langle n^{\prime} q - m_1| j \rangle \right] \nonumber \\
&\quad +\kappa_2 \langle n | \cos \hat{x} | n^{\prime} \rangle
\left[ \langle i | n,q \rangle \langle n^{\prime},q + m_2| j \rangle
+ \langle i | n,q \rangle \langle n^{\prime},q - m_2| j \rangle
\right]\,. \nonumber
\end{align}
Recalling the structure of the states $| \phi^0_{i,Q_i} \rangle$
given in Eq. (\ref{eqn:barredstates}), we see that the sum
\begin{equation}
\sum_q \langle i | n, q \rangle \langle n^{\prime}, q + m | j\rangle
= \sum_q \langle \phi^0_{i,Q_i} | n,q  \rangle \langle n^{\prime}, q
+ m| \phi^0_{j,Q_j} \rangle\,,
\end{equation}
can be non-zero only when $Q_j + m = Q_i + k \, m_0$ with $k \in Z$.
Thus, the only non-zero matrix elements of $\hat{V}$ in Eq.
(\ref{eqn:firstorder_twores}) are
\begin{align}
V_{a,b} &= \frac{\kappa_1}{4} \sum_{n,q} \left( \langle a | n,q
\rangle \langle n+1, q-m_1 | b \rangle  + \langle
a | n,q \rangle \langle n-1,q-m_1 | b \rangle \right) \nonumber\\
&= \frac{\kappa_1}{4} \sum_{n,q} \left(\langle \bar{\phi}^{\,0}_a |
n,q \rangle \langle n+1, q | \bar{\phi}^{\,0}_b \rangle + \langle
\bar{\phi}^{\,0}_a | n,q \rangle \langle n-1,q | \bar{\phi}^{\,0}_b
\rangle \right) \nonumber \\
&= \kappa_1 \, \frac{\langle \langle \bar{\phi}^{\,0}_a | ( \cos \hat{x}
\otimes \mathbb{I}) | \bar{\phi}^{\,0}_b \rangle \rangle}{2}
\label{eqn:matrix_element_Vab}
\end{align}
and
\begin{align}
V_{b,c} &= \frac{\kappa_2}{4} \sum_{n,q} \left( \langle b | n,q
\rangle \langle n+1, q-m_2 | c \rangle  + \langle
b | n,q \rangle \langle n-1,q-m_2 | c \rangle \right) \nonumber \\
&= \frac{\kappa_2}{4} \sum_{n,q} \left( \langle \bar{\phi}^{\,0}_b |
n,q \rangle \langle n+1, q | \bar{\phi}^{\,0}_c \rangle + \langle
\bar{\phi}^{\,0}_b | n,q \rangle \langle n-1,q | \bar{\phi}^{\,0}_c
\rangle \right) \nonumber \\
&= \kappa_2 \, \frac{\langle \langle \bar{\phi}^{\,0}_b | ( \cos \hat{x}
\otimes \mathbb{I}) | \bar{\phi}^{\,0}_c
\rangle \rangle }{2}\label{eqn:matrix_element_Vbc}
\end{align}
where the second and third equalities for each matrix element have
been written in terms of the Floquet states in $\bar{\mathcal{H}}$,
using Eq. (\ref{eqn:barredstates}). Subtracting $\epsilon^0_a
(C_a,C_b,C_c)^{T}$ from both sides, redefining $\epsilon^{\prime} -
\epsilon^0_a \rightarrow \epsilon^{\prime}$, and defining $v_{a,b} =
V_{a,b}/\kappa_1$ and $v_{b,c} = V_{b,c}/\kappa_2$, Eq.
(\ref{eqn:firstorder_twores}) becomes
\begin{equation} \label{eqn:STIRAP_twores}
\left( \begin{array}{ccc}
0 & \lambda \kappa_1 \,v_{a,b} & 0 \\
\lambda \kappa_1 \,v_{b,a} & \Delta_1 & \lambda \kappa_2 \,v_{b,c} \\
0 & \lambda \kappa_2 \,v_{c,b} & \Delta_1-\Delta_2
\end{array} \right)
\left( \begin{array}c
C_a \\
C_b \\
C_c
\end{array} \right)
= \epsilon^{\prime} \left( \begin{array}c
C_a \\
C_b \\
C_c
\end{array} \right)\,,
\end{equation}
which is equivalent to the desired model system in Eq.
(\ref{eqn:STIRAP_detunings}).

\begin{figure}
  \centering
  \includegraphics[width=1\textwidth]{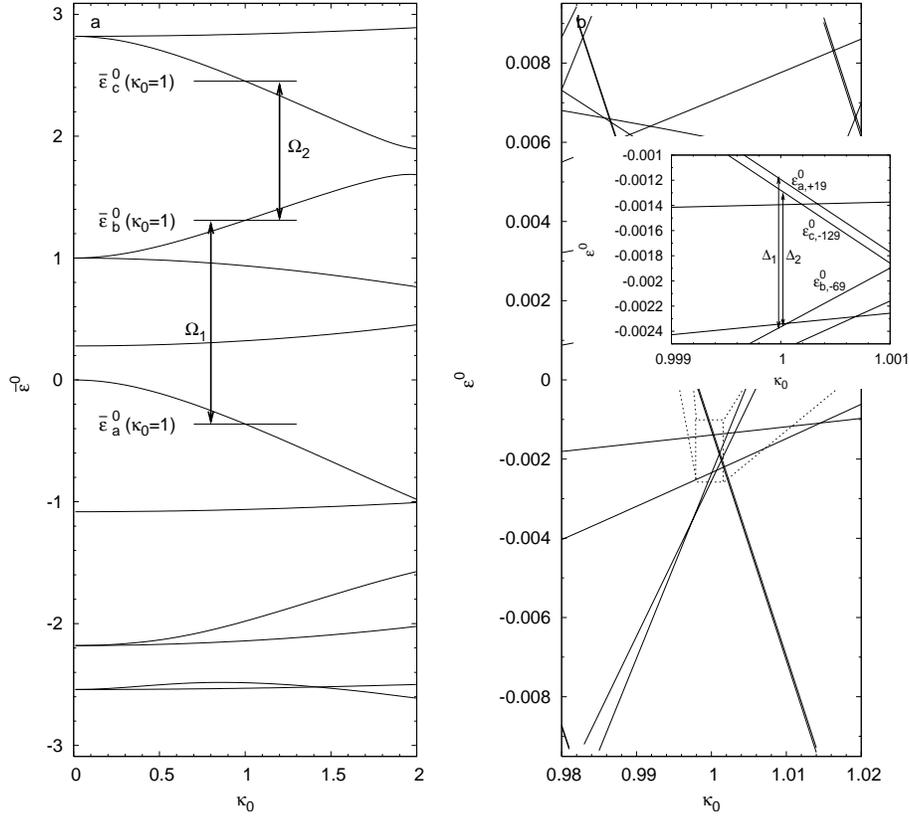}
  \caption{Eleven eigenvalues of the two-resonance Floquet Hamiltonian
    with $\omega_0\approx 6.18$, viewed as an operator in
    $\bar{\mathcal{H}}$, plotted as a function of $\kappa_0$ (a).  The
    corresponding ``unbarred'' Floquet eigenvalues of the same
    Hamiltonian, viewed as an operator in the space $\mathcal{H}$
    with $\omega=\omega_0/325$, are shown in the zone labeled by
    $\epsilon^{\star} =-\omega/2$ (b).  The near-degeneracy of three
    eigenvalues $\epsilon^0_{a,Q_a}$, $\epsilon^0_{b,Q_b}$ and
    $\epsilon^0_{c,Q_c}$ at $\kappa_0=1$ can be seen in the inset. It
    is evident that there are other eigenvalues nearly degenerate with
    these three, however these need not be considered in
    Eq. (\ref{eqn:twores_phi0}) since their respective $Q$-values will
    yield zero-valued matrix elements.}
  \label{fig:twores_stirap}
\end{figure}
\begin{figure}
  \centering
  \includegraphics[width=1\textwidth]{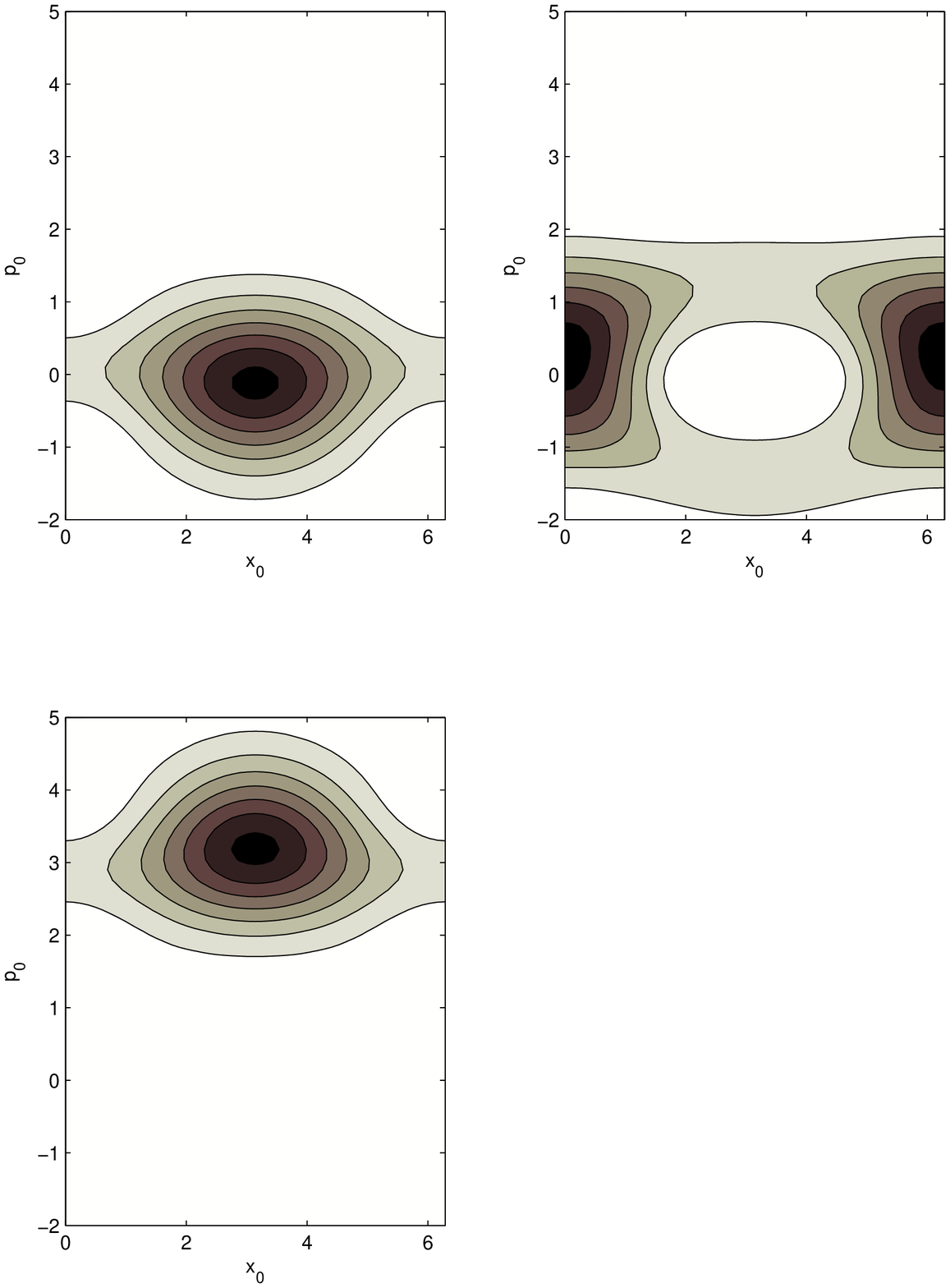}
  \caption{ The Husimi representations of three Floquet states of the
  two-resonance Hamiltonian $\left( |\bar{\phi}^{\,0}_a
  (t)\rangle,\,\, |\bar{\phi}^{\,0}_b (t)\rangle \,\, {\rm and} \,\, |
  \bar{\phi}^{\,0}_c (t)\rangle,\,\,{\rm
  clockwise\,\,from\,\,top\,\,left}\right)$ viewed at time $t=0$.  The
  parameter values are $\kappa_0=1$ and $\omega_0 \approx 6.18$.}
  \label{fig:2res_husimis}
\end{figure}
Again, we provide a example system on which to demonstrate the analysis.
Figure \ref{fig:twores_stirap}a shows some eigenvalues
$\bar{\epsilon}^{\,0}_{\alpha}$ of the two-resonance Floquet Hamiltonian, in
the zone labeled by $\epsilon^{\star}=-\omega_0/2$ with
$\omega_0=6.180339887$, plotted as functions of $\kappa_0$.  A triplet of
eigenvalues $\bar{\epsilon}^{\,0}_a < \bar{\epsilon}^{\,0}_b
<\bar{\epsilon}^{\,0}_c$ has been chosen at $\kappa_0=1$ for STIRAP
coupling.  The Husimi functions of the three corresponding eigenstates are
shown in Figure \ref{fig:2res_husimis}.  The values of these eigenvalues
satisfy $\omega_1 = \bar{\epsilon}^{\,0}_b - \bar{\epsilon}^{\,0}_a =
1.67227495$ and $\omega_2 = \bar{\epsilon}^{\,0}_c - \bar{\epsilon}^{\,0}_b
= 1.14207065$.  Therefore we seek simultaneous rational approximates $(
m_1/m_0,m_2/m_0)$ to the pair $(0.270579771,0.184790913)$. Performing a
numerical exhaustive search, we find that the integer vector $\vec{m} =
(325, 88,60)$ provides detunings $\Delta_1=-1.17 \times 10^{-3}$ and
$\Delta_2 = - 1.08 \times 10^{-3}$, which satisfy the required conditions of
$\Delta_1 \sim O(\lambda) << 1$ and $|\Delta_1 - \Delta_2| << \lambda$.  The
unbarred eigenvalues in the zone $\epsilon^{\star}=-\omega/2 = -\omega_0/(2
\times 325)$ are shown in Figure \ref{fig:twores_stirap}b.  The detunings
$\Delta_1$ and $\Delta_2$ can be seen in the enlarged section of the graph
(inset), separating $\epsilon^0_{b,-69}$ from $\epsilon^0_{a,19}$ and
$\epsilon^0_{c,-129}$, respectively.  The coefficients
\begin{align}
v_{a,b} &= \frac{0.4442}{2}\\
v_{b,c} &= \frac{-0.0673}{2}
\end{align}
are calculated using Eqs. (\ref{eqn:matrix_element_Vab}) and
(\ref{eqn:matrix_element_Vbc}) after numerical diagonalization of
$\hat{H}^0_F$ in $\bar{\mathcal{H}}$.

\begin{figure}
  \centering
  \includegraphics[width=\textwidth]{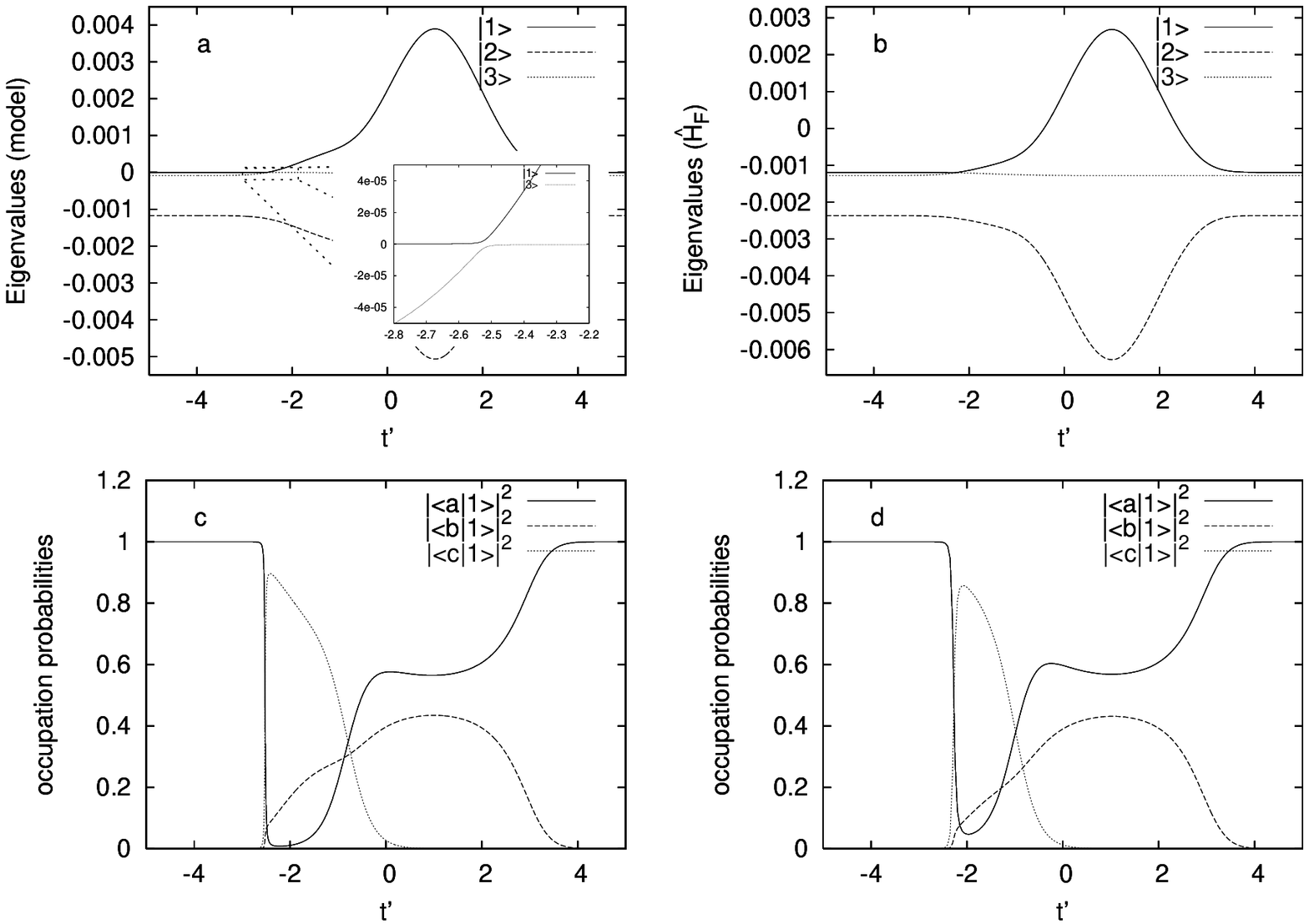}
  \caption{The adiabatic dynamics of the $3$-level model in
  Eq. (\ref{eqn:STIRAP_twores}) (a and c) and the three relevant
  states of the corresponding full two-resonance Floquet system with
  $\omega_0 \approx 6.18$ (b and d) for the example at $\kappa_0=1$.
  The perturbation amplitude functions $\kappa_1(t^{\prime})$ and
  $\kappa_2(t^{\prime})$ are the same as in Figure
  \ref{fig:pend-adiab_3st_n_full}a, and $\lambda =0.02$; model
  parameters are as determined in the text. The eigenvalues under the
  application of the perturbation are show in (a) \& (b). The
  influence of $\kappa_1(t^{\prime})$ on the adiabatic eigenvalues is
  stronger than that of $\kappa_2(t^{\prime})$ because $|v_{a,b}| >
  |v_{b,c}|$. Overlaps of the $| 1 \rangle$ adiabatic eigenvector with
  the unperturbed states are shown in (c) and (d). The inset in (a)
  shows a sharp avoided crossing which prohibits the STIRAP-like
  transition in the adiabatic limit.}
  \label{fig:2res-adiab_3st_n_full}
\end{figure}
The adiabatic dynamics of the model system in Eq.
(\ref{eqn:STIRAP_twores}) and the full system in Eq.
(\ref{eqn:twores_pert_floqham}) are shown in Figure
\ref{fig:2res-adiab_3st_n_full}, using the same parameterization of the
$\kappa_i$ as in Eq. (\ref{eqn:kappai_gauss}) and $\lambda=0.02$.
Although good agreement is seen between the two, it is evident that the
STIRAP transition between eigenstates $|a \rangle$ and $| c \rangle$ is not
achieved in either case. The reason for this failure is a narrow avoided
crossing at $t^{\prime} \approx -2.5$ between the eigenvalues of adiabatic
states $|1 \rangle$ and $|3 \rangle$ (see inset of Figure
\ref{fig:2res-adiab_3st_n_full}a), which affects a transition between
unperturbed states $|a \rangle$ and $|c \rangle$ {\em before} the STIRAP
transition.  This type of avoided crossing, reversing the effects of the
desired transition, will always exist in the adiabatic limit when a matrix
of the type given in Eq. (\ref{eqn:STIRAP_detunings}) is used for STIRAP
evolution because of the non-degeneracy of the eigenvalues of $|a \rangle$
and $|c \rangle$. Although this model does allow for a broad STIRAP-type
transition, the resulting change in character of the adiabatic state $|1
\rangle$ as $t^{\prime}$ passes from $-\infty$ to $\infty$ requires that its
eigenvalue change from $0$ to $\Delta_1 - \Delta_2$ .

The problem with the adiabatic model can be avoided in the numerical
or experimental achievement of a STIRAP transition in one of two
ways. First, it is possible to achieve a non-adiabatic evolution of
the system which is slow enough to guarantee a STIRAP transition, but
too rapid to ``see'' the problematic sharp avoided crossing. Second,
one can abandon time-periodicity and apply resonant coupling fields,
reducing the model to the classic form in Eq. (\ref{eqn:STIRAP}).  The
efficacy of both methods can be seen in Figure
\ref{fig:evolve_2res_w-hus}a, where long-time evolution of both the
model (see Appendix \ref{appendix:b}) and the full Schr\"odinger
equation yields a STIRAP-like transition in the case of resonant and
detuned coupling fields. As in the pendulum case, the resonant
coupling provides a faster approach to the transition.  The evolution
of a state initially prepared in the stationary wave eigenstate $| a
\rangle$ is shown to pass into the traveling wave state $| c \rangle$
under resonant coupling in Figure \ref{fig:evolve_2res_w-hus}b.
\begin{figure}
  \centering
  \includegraphics[width=0.8\textwidth]{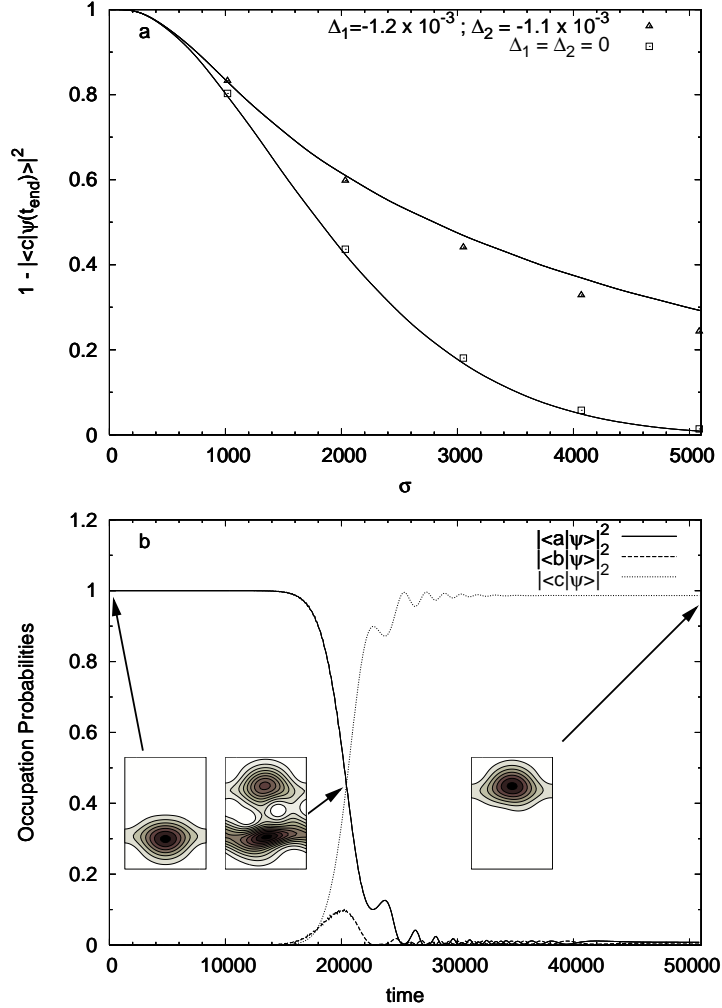}
  \caption{The numerical evolution of a state initially localized in
the stationary cosine wave of the two-resonanace system ($|\langle
\phi^0_a(0)| \psi(0) \rangle|^2 = 1$), under the influence of
$\hat{V}(t)$ for the example described in the text. The probability of
a transition to any state other than $| c \rangle$ is plotted versus
the width of the coupling fields $\sigma=\sigma_1=\sigma_2$ (a), with
evolution under resonant coupling ($\Omega_1=\omega_1$ and
$\Omega_2=\omega_2$) plotted with triangles and evolution under
near-resonant coupling (with $\Delta_1$ and $\Delta_2$ as shown in
Figure \ref{fig:twores_stirap}) plotted with squares. Solid lines are the
corresponding values predicted by the numerical evolution of the
three-state model.  Figure (b) shows the evolution of the
occupation probabilities for the case of $\sigma \approx 5000$ and
resonant coupling (axes on inset Husimi functions, each taken at an
integer multiple of the period $\bar{T}$, are the same as in
Fig \ref{fig:2res_husimis}).} 
\label{fig:evolve_2res_w-hus}
\end{figure}

Using the non-dimensionalization presented at the end of Appendix
\ref{appendix:a}, we can return to dimensional variables and determine
the experimental conditions necessary for a STIRAP transition of the
type described here.  Consider a system of cold cesium atoms
interacting with a system of lasers tuned near the $\rm{D}_2$
transition (as in References \cite{raizen_science,steck2002}),
yielding a recoil frequency of $\omega_r \approx 1.3 \times 10^4$
Hz. For the example considered above, the traveling cosine wave is
therefore generated by counterpropagating lasers with frequencies
offset by $\delta \omega /2\pi \approx 10$ kHz; amplitude modulation
frequencies corresponding to $\Omega_1/2\pi$ and $ \Omega_2/2\pi$ are
$11$ and $7.7$ kHz, respectively. The relationship between
dimensionless time $t$ and physical time $t_{phys}$ for this system is
\begin{equation}
t_{phys} \approx 1.0 \times 10^{-5} \, t\,.
\end{equation}
Therefore, the near-100\% transition between the stationary and
travelling lattices shown in Figure \ref{fig:evolve_2res_w-hus}b would
require half a second in the laboratory.  The approach to adiabatic
evolution can be achieved more rapidly by increasing the coupling
strength $\lambda$ \cite{kuklinski1989}, as long as the assumption of
small perturbation ($\lambda << \kappa_0$) remains valid.  In
numerical experiments, we were able to decrease the transfer time for
the preceding example (with resonant coupling) by a factor of five,
while maintaining 90\% efficiency, by setting $\lambda \approx 0.1$.
Selection of a larger value of $\kappa_0$, i.e. deeper wells in the
optical lattice, would allow for a larger value of $\lambda$ and
shorter transfer times.

\section{\label{sec:4} Conclusions}
We have demonstrated a method for the coherent transfer of ultracold
atoms from the well of a stationary optical lattice into that of a
travelling lattice.  The effective Hamiltonian for an atom in such a
lattice, constructed by adiabatic elimination of the internal
electronic structure, provides a system of eigenstates which determine
the center-of-mass dynamics of the atom.  We have shown that small,
harmonic modulations of the lattice amplitude can allow for a
STIRAP-type subsystem of the Schr\"odinger equation for this effective
Hamiltonian, with which transitions between these eigenstates can be
induced.

\section{Acknowledgements}
The authors thank the Robert A. Welch Foundation (Grant No. F-1051)
and the Engineering Research Program of the Office of Basic Energy
Sciences at the U.S. Department of Energy (Grant No.
DE-FG03-94ER14465) for support of this work.  They also thank the
Texas Advanced Computing Center (TACC) for the use of their
computing facilities in performing calculations for this paper.

\appendix
\section{\label{appendix:a}Experimental creation of the effective Hamiltonian}

In this appendix, we show how the effective Hamiltonian in Equation
(\ref{eqn:tworesham}) can be used to describe an experimental system
of lasers impinging on non-interacting alkali atoms.  This analysis
involves the consideration of a two-level subsystem of the atom's
electronic levels, application of the rotating-wave approximation,
and adiabatic elimination of the excited level, to obtain a
spatially and temporally-periodic potential for atoms in the ground
state.

We begin by considering the Hamiltonian of this two level system, in dipole
interaction with a $z$-polarized electric field:
\begin{equation}
H = H_{atom} + H_{int}\,,
\end{equation}
with
\begin{equation}
H_{atom} = \hbar \omega_{at} | e \rangle \langle e | + \frac{p_x^2}{2
m} \left( |e \rangle \langle e | + | g \rangle \langle g| \right) \,,
\end{equation}
and
\begin{equation}
H_{int}= - d E_z(x,t) \left( |e
\rangle \langle g| + |g \rangle \langle e| \right)\,,
\end{equation}
where $\hbar \omega_{at}$ is the energy spacing of the two levels,
$p_x$ is the atomic momentum operator in the $x$-direction, and $d
\equiv \langle e | \hat{d}_z | g\rangle = \langle g | \hat{d}_z |e
\rangle$ is the dipole matrix element coupling the ground state $| g
\rangle$ to the excited state $| e \rangle$. The total electric
field amplitude $E_z(x,t)$ is assumed to be the superposition of the
electric fields due to $N$ lasers, all polarized along the $z$
direction, so that
\begin{equation}
E_z(x,t) = \sum_{j=1}^N E^{(j)} \cos \left[ (k_L + \frac{\delta k_j}{2})x + \sigma_j
  (\omega_L + \frac{\delta \omega_j}{2})t + \phi_j \right]
\end{equation}
where $E^{(j)}$ is the amplitude of the $j$th laser, $\omega_L$ is a
positive reference frequency and $k_L$ its corresponding wavevector,
$\sigma_j$ can be $\pm 1$, and $\delta k_j = \delta \omega_j/c$ (the
usefulness of this form will be evident below).  We can then write
\begin{equation} \label{eqn:Efield}
E_z(x,t) = A(x,t)\, {\rm e}^{- i \omega_L t} + A^{\star}(x,t)\, {\rm e}^{i
  \omega_L t} \,,
\end{equation}
with
\begin{equation}
A(x,t) = \sum_j \frac{E^{(j)}}{2} \exp\left\{ - i \sigma_j \left[(k_L
+ \frac{\delta k_j}{2})x + \sigma_j \frac{\delta \omega_j}{2}t +
\phi_j \right] \right\}\,.
\end{equation}
Under a time-dependent unitary transformation of the Schr\"odinger equation,
the Hamiltonian transforms like
\begin{equation}
H \rightarrow U\,H\,U^{\dagger} + i \hbar \frac{\partial U}{\partial
  t}\, U^{\dagger}\,.
\end{equation}
Using the unitary matrix
\begin{equation}
U = \exp\left[ i \omega_L \, | e \rangle \langle e| \, t \right]\,,
\end{equation}
to transform to the rotating frame of the laser leaves the
Hamiltonian as
\begin{multline}
H = \hbar \Delta |e\rangle \langle e | + \frac{p_x^2}{2 m} (|e
\rangle \langle e | + |g \rangle \langle g | ) \\ - d \, E_z(x,t)
\left( |e \rangle \langle g| {\rm e}^{i \omega_L t}+ | g \rangle
\langle e | {\rm e}^{-i \omega_L t}\right)\,,
\end{multline}
where $\Delta = \omega_{at} - \omega_L$ is the detuning of the
reference laser frequency from the atomic transition.

Let us now make the rotating wave approximation by inserting the
form of $E_z$ in Eq. (\ref{eqn:Efield}) and neglecting terms with
high-frequency exponential dependence (i.e. ${\rm e}^{\pm i 2
\omega_L t}$). The Hamiltonian then takes the form
\begin{equation}
H = \hbar \Delta | e \rangle \langle e| + \frac{p_x^2}{2 m} ( |e
\rangle \langle e| + |g \rangle \langle g|)- d \left( A(x,t) |e
\rangle \langle g| + A^{\star}(x,t) |g \rangle \langle e | \right)\,.
\end{equation}
Writing an arbitrary state $| \psi \rangle = \psi_g(x,t) |g\rangle +
\psi_e(x,t) | e \rangle$, the Schr\"odinger equation can be written
\begin{align}
i \hbar \frac{\partial \psi_g}{\partial t} &= -\frac{\hbar^2}{2 m}
\frac{\partial^2}{\partial x^2} \psi_g - d A^{\star}(x,t) \psi_e \\ i
\hbar \frac{\partial \psi_e}{\partial t} &= -d A(x,t) \psi_g + \left(
\hbar \Delta - \frac{\hbar^2}{2 m} \frac{\partial^2}{\partial x^2}
\right) \psi_e \,.
\end{align}
Adiabatic elimination of the excited state is performed by assuming
that the detuning of the laser $\Delta$ is large enough to allow us
to neglect the time and space derivatives of the excited state.
Thus, atoms prepared in the ground state will remain there and we
are left with an {\em effective Hamiltonian} for their evolution:
\begin{equation}
i \hbar \frac{\partial \psi_g}{\partial t} = H_{eff} \, \psi_g
\quad;\quad H_{eff} = \frac{p_x^2}{2m} - \frac{d^2 |A(x,t)|^2}{\hbar
  \Delta}\,.
\end{equation}

The particular form of $A(x,t)$ will depend on the choice of lasers.
A pair of counter-propagating lasers with equal carrier frequencies
($E^{(1)}=E^{(2)}\equiv E$; $\delta\omega_1= \delta\omega_2 = 0$;
$\sigma_1 = -\sigma_2$; $\phi_1=\phi_2=0$) will produce a
time-independent, periodic potential, i.e
\begin{equation}
A_{stand}(x,t) = \frac{E}{2} \left( {\rm e}^{i k_L x} + {\rm e}^{-i
k_L x} \right) \quad \rightarrow \quad |A_{stand}(x,t)|^2 \sim
\frac{E^2}{2} \cos (2 k_L x) \,,
\end{equation}
where we have neglected constant terms.  Similarly, two
counter-propagating lasers with slightly offset frequencies
($E^{(1)}=E^{(2)}\equiv E$ ; $\delta\omega_1 = - \delta\omega_2 \equiv
\delta \omega$ ; $\sigma_1 = -\sigma_2$; $\phi_1=\phi_2=0$) will
produce a travelling periodic potential:
\begin{multline}
A_{trav}(x,t) = \frac{E}{2} \left\{ {\rm e}^{i \left[(k_L +
  \frac{\delta k}{2})x - \frac{\delta \omega}{2} t \right]} + {\rm
  e}^{- i \left[(k_L - \frac{\delta k}{2})x - \frac{\delta \omega}{2}
  t \right]} \right\} \\ \rightarrow \quad |A_{trav}(x,t)|^2 \sim
  \frac{E^2}{2} \cos (2 k_L x - \delta \omega t) \,.
\end{multline}

If we combine these two pairs of lasers, we create an effective
potential with the desired terms of Eq. (\ref{eqn:tworesham}),
namely
\begin{equation}\label{eqn:tworesham_dimensional}
H = \frac{p_x^2}{2m} - \frac{d^2 E^2}{2 \hbar \Delta} \left[ \cos (2
k_L x) + \cos (2 k_L x - \delta \omega \, t) \right] \,.
\end{equation}
It is clear, however, that the $|A(x,t)|^2$ for such a system will
also contain unwanted cross-terms, which we have neglected in
writing Eq. (\ref{eqn:tworesham_dimensional}). In order to minimize
the effect of these cross terms, we offset the carrier frequency of
the second pair by some amount $ \Delta \omega$ ($E^{(1)}=E^{(2)}=
E^{(3)} = E^{(4)}\equiv E$ ; $\delta\omega_1 = \delta\omega_2 = 0$ ;
$\delta\omega_3 = \Delta \omega + \delta \omega$ ; $\delta\omega_4 =
\Delta \omega - \delta \omega$ ; $\sigma_1 = \sigma_3 = -\sigma_2 =
- \sigma_4$; and $\phi_i=0 \; \forall \,i$), where $\omega_L >>
\Delta \omega >> \delta \omega$. This yields,
\begin{multline} \label{eqn:Atwores}
|A_{two-res}(x,t)|^2 \sim \frac{E^2}{2} \left[ \cos (2 k_L x) + \cos (2
 k_L x - \delta \omega t) \right. \\
\left. + \cos (\Delta k x) \cos (\Delta \omega t) + \cos (2 k_L x) \cos
 (\Delta \omega t) \right] \,,
\end{multline}
where $\Delta k = \Delta \omega/c$ and we have retained only the
highest-order terms in the frequencies and wavevectors (e.g. $\delta
\omega$ is neglected in the presence of $\Delta \omega$).  The last
two terms in this equation present high-frequency oscillations,
depending on the particular value of $\Delta \omega$.  As a concrete
example, we can consider a system of cesium atoms.  In references
\cite{raizen_science,steck2002}, the laser light was detuned by
$\Delta \sim 10^{11}$Hz from the ${\rm D}_2$ line ($\omega_L \sim
10^{15}$Hz) and a modulation of $\delta \omega \sim 10^5$Hz was
applied to the standing lattice to affect travelling terms in the
effective potential.  Therefore, an offset of the carrier frequency
for the second pair of lasers in the hundreds of MHz will satisfy
$\Delta >> \Delta \omega >> \delta \omega$, and allow one to safely
neglect the last two terms in the square brackets of
Eq. (\ref{eqn:Atwores}) \cite{raizen_talk}.

In order to obtain the Hamiltonian in Eq. (\ref{eqn:tworesham}), we
change to dimensionless variables ($p^{\prime}$, $x^{\prime}$,
$H^{\prime}$, $t^{\prime}$, $\omega^{\prime}$) as follows.  Let
$p^{\prime}=\frac{p_x}{2 \hbar k_L}$, $x^{\prime}=2 k_L x$,
$H^{\prime}=\frac{1}{4 \hbar \omega_r} H$, $t^{\prime}=8 \omega_r t$,
and $\omega^{\prime} = \frac{1}{8 \omega_r} \delta \omega$ where the
recoil frequency of an atom is $\omega_r = \frac{\hbar k_L^2}{2 m}$.
The Hamiltonian in Eq.  (\ref{eqn:tworesham_dimensional}) then takes
the form
\begin{equation}
H^{\prime} = \left(p^{\prime}\right)^2 + \kappa \left[ \cos
\left(x^{\prime} \right) + \cos \left(x^{\prime} + \omega^{\prime} t
\right) \right]
\end{equation}
where $\kappa \equiv - \frac{d^2 E^2}{8 \omega_r \hbar^2
\Delta}$. Removing the primes, we obtain the desired Hamiltonian.  It
is important to note that in these dimensionless units, changes in
momentum due to the interaction of an atom with the lasers are
integer-valued.  Moreover, experimental techniques allow for
the preparation of atoms in a very narrow range of momentum values
about zero \cite{raizen_science,steck2002}.  Therefore, the
eigenvalues of the momentum operator will take only integer values,
i.e. $\hat{p} | n \rangle = n | n \rangle$ with $n \in \mathbb{Z}$.

\section{\label{appendix:b} ``Evolution'' of a Floquet Hamiltonian}
In this appendix, the $(t, t^{\prime})$ formalism due to Peskin and
Moiseyev \cite{peskin_moiseyev1993,fleischer_moiseyev2005} is used to
justify the time-parametrization of $\kappa_1$ and $\kappa_2$ in the
model Hamiltonians in Eqs. (\ref{eqn:STIRAP_pend}) and
(\ref{eqn:STIRAP_twores}).  These models are each subsystems of a {\em
Floquet} Hamiltonian which was constructed under the assumptions that
$\kappa_1$ and $\kappa_2$ were constant and the Schr\"odinger equation
was time-periodic.  The subsequent parametrization of such a system by
non-periodic functions of time therefore requires a more rigorous
explanation.  Here, we show that a physical system represented by a
Hamiltonian with both periodic and arbitrary dependence on time, can
be associated to Floquet-like Hamiltonian in an extended Hilbert space
where the periodic time-dependence has been reduced to dependence on a
coordinate.  This Hamiltonian is termed ``Floquet-like'' because its
dependence on the {\em coordinate} time is identical to a Floquet
Hamiltonian's dependence on time.  The remaining arbitrary
time-dependence of the Floquet-like Hamiltonian determines, via the
Schr\"odinger equation, a dynamics in the extended space from which
the dynamics of the original system can be recovered.

Consider the Schr\"odinger equation for a time-dependent Hamiltonian
\begin{equation} \label{eqn:tdse}
i  \frac{\partial}{\partial t} \psi(x;t) = H(x;t) \psi(x;t) \,,
\end{equation}
where $x$ can be considered a single spatial coordinate or a set of
coordinates and $\hbar$ has been set to unity by non-dimensionalization of
the variables.  We will associate to $H(x;t)$ a Hamiltonian of one more
coordinate $H_F(x,t^{\prime};t)$ which is a Hermitian operator in a larger
Hilbert space, extended to include this new coordinate $t^{\prime}$. The
relationship between the two Hamiltonians is defined by
\begin{equation} \label{eqn:HF_ttprime}
H_F(x,t^{\prime};t) = \bar{H}(x,t^{\prime};t) - i 
\frac{\partial}{\partial t^{\prime}}\,,
\end{equation}
with
\begin{equation} \label{eqn:Hbar}
\bar{H}(x,t^{\prime};t) \vert_{t^{\prime}=t} = H(x;t) \,.
\end{equation}
Clearly, $H(x;t)$ does not uniquely determine $H_F(x,t^{\prime};t)$.  The
time-evolution of a state $\bar{\psi}(x,t^{\prime};t)$ in the extended space
is governed by the Schr\"odinger equation
\begin{equation} \label{eqn:HF_extendedschrod}
i  \frac{\partial}{\partial t} \bar{\psi}(x,t^{\prime};t) =
H_F(x,t^{\prime};t) \bar{\psi}(x,t^{\prime};t)\,,
\end{equation} 
which can also be written
\begin{equation}
i  \left[ \left( \frac{\partial}{\partial t} + \frac{\partial}{\partial
t^{\prime}} \right) \bar{\psi}(x,t^{\prime};t) \right] =
\bar{H}(x,t^{\prime};t) \bar{\psi}(x,t^{\prime};t) \,.
\end{equation}
If we take this equation at the cut $t^{\prime}=t$, it becomes
\begin{equation} \label{eqn:barred_tcut}
i  \frac{\partial}{\partial t} \left[ \bar{\psi}(x,t^{\prime};t)
\vert_{t^{\prime}=t} \right] = H(x;t) \left[ \bar{\psi}(x,t^{\prime};t)
\vert_{t^{\prime}=t} \right] \,,
\end{equation}
where we have used the identity
\begin{equation}
\frac{\partial}{\partial t} \left[ \bar{\psi}(x,t^{\prime};t)
\vert_{t^{\prime}=t} \right] = \left[\left( \frac{\partial}{\partial t} 
+ \frac{\partial}{\partial t^{\prime}} \right) 
\bar{\psi}(x,t^{\prime};t) \right]_{t^{\prime}=t} \,.
\end{equation}
Comparing Eqs. (\ref{eqn:barred_tcut}) and (\ref{eqn:tdse}), we see that
the evolution of a state in the original system can be determined by
evolution in the extended system using 
\begin{equation} \label{eqn:psi_psibar}
\psi(x;t) = \bar{\psi}(x,t^{\prime};t) \vert_{t^{\prime}=t}\,,
\end{equation}
and provided the same initial condition
\begin{equation} \label{eqn:psi_psibar_t0}
\bar{\psi}(x,t^{\prime};t) \vert_{t^{\prime}=t=0} = \psi(x,0) 
\end{equation}
is used in each space.

We now apply this formalism to STIRAP transitions in the two-resonance
Hamiltonian.  The evolution plotted in Figure \ref{fig:evolve_2res_w-hus}b
was performed by numerical integration of the Schr\"odinger equation, using
the Hamiltonian
\begin{multline}
H(x;t) = - \frac{\partial^2}{\partial x^2} + \kappa_0
\left[ \cos x + \cos ( x - \omega_0 t) \right] \\
+ \lambda \cos x \left[ \kappa_1(t) \cos (\Omega_1 t) + \kappa_2(t)
\cos (\Omega_2 t) \right] \,,
\end{multline}
where $\kappa_1$ and $\kappa_2$ were given Gaussian time-depenence in order
to affect the STIRAP-like transition.  The solid lines plotted in Figure
\ref{fig:evolve_2res_w-hus}a, were determined by evolution of a
Schr\"odinger equation using the time-parametrized three-level model in Eq.
(\ref{eqn:STIRAP_twores}).  Using the above analysis we can show that,
modulo the perturbation theory approximations, these two methods of
time-evolution are equivalent.  We define a Hamiltonian in the extended space
\begin{multline}
\bar{H}(x,t^{\prime};t) = - \frac{\partial^2}{\partial x^2} + \kappa_0
\left[ \cos x + \cos ( x - \omega_0 t^{\prime}) \right] \\
+ \lambda \cos x \left[ \kappa_1(t) \cos (\Omega_1 t^{\prime}) + \kappa_2(t)
\cos (\Omega_2 t^{\prime}) \right] \,,
\end{multline}
which satisfies Eq. (\ref{eqn:Hbar}) for the two-resonance Hamiltonian
and has the property that functions periodic in time are now functions
of the extra coordinate, while the amplitudes of the modulations are
functions of the usual time parameter.  The full Hamiltonian in the
extended space $H_F(x,t^{\prime};t)$, defined by
Eq. (\ref{eqn:HF_ttprime}), has the same dependence on $t^{\prime}$
that the Floquet Hamiltonian in Eq.  (\ref{eqn:twores_pert_floqham})
has on time $t$.  Therefore, the entire perturbation analysis
performed on the Floquet Hamiltonian in Section \ref{sec:3} would
proceed in identical fashion on $H_F(x,t^{\prime};t)$, yielding a
time-parametrized three-level model. If Eqs. (\ref{eqn:psi_psibar})
and (\ref{eqn:psi_psibar_t0}) are satisfied, the ``time-parametrized''
Floquet Hamiltonian can be used to determine the physical evolution.

\pagebreak

\end{document}